\documentclass[a4paper, 10pt]{article}
\usepackage[left= 0.5in, right= 0.5in, top=1in, bottom = 1 in]{geometry}
\usepackage{amsfonts}
\usepackage{amsmath}
\usepackage{amssymb}
\usepackage{graphicx}
\usepackage{bm}
\usepackage{mathrsfs}
\usepackage{caption}
\usepackage{subcaption}
\usepackage{color}
\usepackage{caption}
\usepackage{natbib}
\usepackage{hyperref}
\usepackage{chngcntr}
\usepackage{gensymb}
\usepackage{tikz}
\usepackage{setspace}
\usepackage[toc,page]{appendix}
\usepackage{cleveref}
\usepackage{authblk}
\usepackage{cancel}
\hypersetup{hyperindex, linkcolor=blue, colorlinks =true, citecolor=blue, urlcolor=magenta}

\allowdisplaybreaks 

\date{ }
	\title{Growth induced instabilities in a  circular hyperelastic plate}

\newcommand{\mbf}[1]{\mathbf{#1}}

\newcommand{\A}{\mathbf{A}}

\newcommand{\changes}[1]{\textcolor{black}{#1}}
\newcommand{\changesg}[1]{\textcolor{black}{#1}}

\author[1]{{Sumit Mehta}}
\author[1]{{Gangadharan Raju}}
\author[2]{{Prashant Saxena}\thanks{Corresponding author email: prashant.saxena@glasgow.ac.uk}}

\affil[1]{Department of Mechanical and Aerospace Engineering \protect\\
Indian Institute of Technology Hyderabad, India}
\affil[2]{Glasgow Computational Engineering Centre, James Watt School of Engineering, \protect\\ University of Glasgow, Glasgow G12 8LT, UK}

\begin{document}
\numberwithin{equation}{section}
\maketitle

\begin{abstract}
In this work, we have explored growth-induced mechanical instability in an isotropic  circular hyperelastic plate.
Consistent two-dimensional governing equations for a plate under a general finite strain are derived using a variational approach.
\changesg{The derived plate equations are}  solved using the compound matrix method for two cases of axisymmetric growth conditions -- purely radial, and combined radial and circumferential growth.
\changes{The effect of growth on the buckling behaviour of the plate (in particular, the  critical growth factor and the associated buckling mode shapes) is investigated for different thickness values. }
These results are applicable to model growth induced deformation in planar soft tissues such as skin.
\end{abstract}

\noindent \textbf{\textit{Note}}: This is the author-generated version of paper to be published in the International Journal of Solids and Structures (2021).

\section{Introduction}
Mechanical instabilities are ubiquitous in nature and often result in pattern formation in thin elastic structures. 
Classical plate theories like Kirchhoff-Love theory, F\"oppl-von K\'arm\'an theory, {and Mindlin-Reissner theory} have been widely used to study the instability behaviour of thin elastic structures \citep{coman2006localized, coman2015asymptotic, li2010buckling}. 
These theories are based on apriori kinematic assumptions which are suitable for solving small strain problems. 
\changesg{Also, these theories when applied to plates under general loading conditions give inconsistent results due to the underlying assumptions of displacement variation along thickness of the plate.}
To overcome the inconsistencies in the classical plate theories, \cite{kienzler2002consistent} developed the consistent plate theory using uniform approximation of unknown variables based on linear elasticity. 
{The} consistent plate theory does not incorporate apriori kinematic assumptions and all the coefficients are treated as independent unknown variables. 
However, small strain theories based on linear elasticity principles are not suitable for finite strain problems. 
To alleviate these problems, a consistent finite-strain plate theory was proposed by \cite{dai2014consistent}. 
This approach was based on the principle of minimisation of potential energy under general three-dimensional loading conditions. 
They derived the two-dimensional plate vector equation by employing variational principle and series expansion of the independent variables about the bottom surface of a hyperelastic plate.
\cite{wang2016consistent} extended this approach to incompressible hyperelastic materials with extra unknown variables to accommodate the incompressibility constraint. 
Mechanical instability is also a common phenomenon in morphoelastic structures \changes{\citep{amar2011new}} and soft biological tissues \changes{\citep{cao2012biomechanical, wu2015growth}}, which exhibit non-linear mechanical response due to growth.

Growth not only changes the mass and geometry of structures but can also alter their mechanical properties and stress state \citep{goriely2017mathematics}.
Growth can induce residual stresses inside the body which result in large deformations leading to instabilities such as wrinkling, folding, creasing, and crumpling \citep{li2012mechanics}. Residual stresses are self equilibrating \changes{ \citep{hoger1986determination,amar2005growth}} generally arising due to the incompatibility of growth. \changes{\cite{goriely2005differential}} and
\cite{vandiver2009differential} studied the buckling of cylindrical elastic structures subjected to differential growth and residual stresses.
\cite{dervaux2009morphogenesis} discussed the nonlinear behaviour of thin elastic structures subjected to growth by considering F\"oppl-von K\'arm\'an plate theory. \cite{moulton2011circumferential} investigated the circumferential instability in differentially growing cylindrical elastic tube subjected to uniform pressure and evaluated the critical pressure for buckling using an incremental theory.
\cite{papastavrou2013mechanics} investigated
\changesg{wrinkling in growing surface by considering membrane with zero thickness in the derivation of potential energy.}
\changes{\cite{wu2015modelling}} performed the bifurcation analysis of uni-directional growing disk reinforced with fibres and determined the effect of fibre anisotropy on critical growth factor. 
Recently, \cite{wang2018consistent} derived a consistent finite-strain plate theory for growth-induced large deformations and investigated the buckling and post-buckling behaviour of a thin rectangular hyperelastic plate under axial growth.

Consistent plate theories with finite-strain have a wide range of applications as they incorporate bending as well as stretching. They are suitable to approximate the behaviour of soft biological tissue such as skin \changes{\citep{tepole2011growing}} and their bifurcation properties under residual stress \changes{\citep{swain2015interfacial,swain2016mechanics}}. \changes{Skin undulation near the edges of wound has been observed in healing experiments on mice \citep{nassar2012calpain,wang2013mouse}}. Beyond biomedical applications, the mechanics of instability with large deformations have important implications on wrinkling analysis of gossamer space structures \citep{wang2009new, deng2019wrinkling} \changes{when exposed to temperature gradients}. Since the last decade, research in understanding buckling/wrinkling instabilities during micro-fabrication in the field of flexible/stretchable electronics  \changes{such as sensory skins used in robotics and wearable communication devices} \citep{rogers2010materials, wei2014fabrication} has increased.

In this manuscript, we have used the consistent finite-strain plate theory given by \cite{wang2018consistent} to derive the governing differential equations for  circular isotropic hyperelastic plates. 
\changes{The symmetry of circular geometry allows us to transform the resulting partial differential equations (PDEs) to ordinary differential equations (ODEs) while still retaining the key aspects of mechanics.
 Our current focus is to comprehend the underlying mechanics of such systems. The formulation is general and can be applied to other geometries by solving the resulting PDEs using numerical techniques such as finite element method.}
Following the multiplicative decomposition approach proposed by \cite{rodriguez1994stress}, the deformation gradient tensor is decomposed into growth tensor, describing change of mass or growth laws and elastic deformation tensor that ensures compatibility (no overlaps) and integrity (no cavitation) \changes{\citep{goriely2007definition}}.
The principle of minimum total potential energy is applied to derive the 3-D governing partial differential equations in the polar coordinate system.
These equations are reduced to two dimensions using a series expansion of unknown functions in terms of the thickness variable.
After establishing the governing equations, we study the instability behaviour of neo-Hookean circular plates growing in radial as well as combined (radial and circumferential) direction. 
The traction conditions at the bottom surface of plate is applied through a Winkler support in both the cases. 
The resulting system of ODEs is stiff and standard shooting methods are not able to evaluate the bifurcation results accurately. 
To resolve this problem, we have used the compound matrix method \citep{ng1979numerical, ng1985compound} to determine the critical point of buckling. 
%

The remainder of this paper is organised as follows.
In \hyperref[section_2]{Section \ref{section_2}}, a general formulation for three-dimensional circular plate is established using a variational formulation. 
The two-dimensional plate governing equations are derived by eliminating the dependence on $Z$ variable using Taylor's expansion. 
In \hyperref[section_4]{Section \ref{section_4}}, we discuss two examples of growth-induced instability in incompressible neo-Hookean circular plate. A detailed discussion of numerical results is provided along with the comparison of existing analytical results for rectangular plate. Finally, we present our conclusions in \hyperref[section_5]{Section \ref{section_5}}.   
 
\subsection{Notation used in this manuscript}

Brackets: Three types of brackets are used. Round brackets ( ) are used to define the functions applied on parameters or variables. Square brackets [ ] are used to clarify the order of operations in an algebraic expression. Curly brackets \{ \} are used to define a set. Square brackets are also used for matrices and tensors. At some places we use the square bracket to define the functional.

\noindent Symbols: A variable typeset in a normal weight font represents a scalar. A lower-case bold weight fonts denotes a
vector and bold weight upper-case denotes the tensor or matrices. 
Tensor product of two vectors $\mbf{a}$ and $\mbf{b}$ is defined as $[\mbf{a} \otimes \mbf{b}]_{ij} = [\mbf{a}]_i [\mbf{b}]_j$.
Tensor product of two second order tensors $\mbf{A}$ and $\mbf{B}$ is defined as either $[\mbf{A} \otimes \mbf{B}]_{ijkl} = [\mbf{A}]_{ij} [\mbf{B}]_{kl} $ or $[\mbf{A} \boxtimes \mbf{B}]_{ijkl} = [\mbf{A}]_{ik} [\mbf{B}]_{jl} $. 
{Higher order tensors are written in bold calligraphic font with a superscript as $ \pmb{\mathcal{A}}^{(i)}$, where superscript `$i$' tells that the function is differentiated $i+1$ times. For example, $\pmb{\mathcal{A}^{(1)}} = \frac{\partial f(\mathbf{A})}{ \partial \mathbf{A} \partial \mathbf{A}} $ is a fourth order tensor. Operation of a fourth order tensor on a second order tensor is denoted as $ [\pmb{\mathcal{A}^{(1)}} : \mbf{A}]_{ij} = [\pmb{\mathcal{A}^{(1)}}]_{ijkl} [\mathbf{A}]_{kl}$ }
Inner product is defined as $\mbf{a} \cdot \mbf{b} = [\mbf{a}]_i [\mbf{b}]_i$ and $\mbf{A} \cdot \mbf{B} = [\mbf{A}]_{ij} [\mbf{B}]_{ij}$.
The symbol $\nabla$ denotes the two-dimensional 
gradient operator. We use the word `Div' to denote divergence in three dimensions.

\noindent Functions: $\det(\mathbf{A})$ denote the determinant of the tensor $\mathbf{A}$. $\text{tr}(\mathbf{A})$ denote the trace of a tensor $\mathbf{A}$. $\text{diag}(a, b, c)$ denotes a second order tensor with only diagonal entries $a, b~ \text{and}~ c$. 
 
\section{Governing equations for growing circular plate  \label{section_2}} 
\begin{figure}
\centering
\includegraphics[scale=0.41]{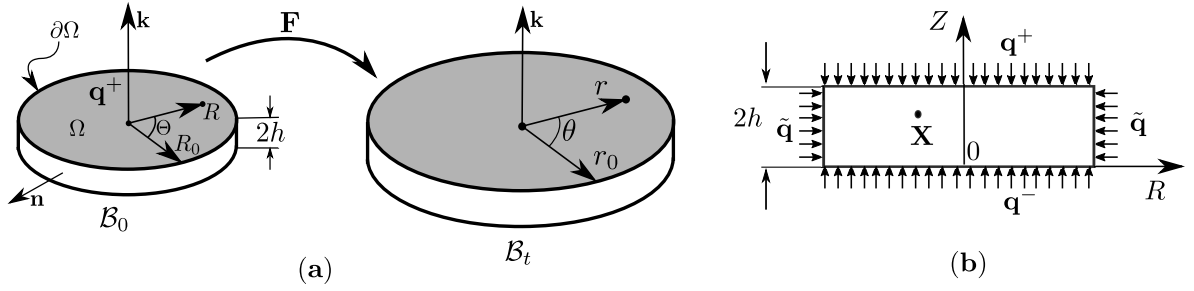}
\caption{Schematic of a circular plate under growth (a) Finite deformation due to growth (perspective) (b) Front view of the plate in reference configuration with applied traction on the top, bottom, and side surfaces}. \label{fig1}
\end{figure}

Consider a thin circular plate with constant thickness $2h$ that occupies the region $ \Omega \times [0,2h]$ in the reference configuration $\mathcal{B}_0 \in \mathscr{R}^3$ and deforms to the current configuration  $\mathcal{B}_t \in \mathscr{R}^3$ as shown in \hyperref[fig1]{Figure~\ref{fig1}}. 
Coordinates of a point in the reference configuration are given by $R, \Theta$ and $Z$ and in the deformed configuration by $r$, $\theta$ and $z$. 
Radius of the plate in the reference configuration is $R_0$. 
Position vectors in the configurations $\mathcal{B}_0$ and $\mathcal{B}_t$ are denoted as $\mathbf{X}({R},~ \Theta,  ~ Z)$ and $\mathbf{x}({r},~ \theta, ~z)$, respectively. The deformation gradient for a circular plate is defined as \citep{dai2014consistent}
\begin{subequations}
\begin{align}
\mathbf{F} &=\frac{\partial \mathbf{x}}{\partial \mathbf{X}}=\frac{\partial \mathbf{x}}{\partial {R}} \otimes \mathbf{e}_R+\frac{\partial \mathbf{x}}{\partial \Theta}\otimes \mathbf{e}_\Theta + \frac{\partial \mathbf{x}}{\partial Z}\otimes \mathbf{k}, \label{2_1}\\
&=\frac{\partial \mathbf{x}}{\partial \mathbf{\zeta}} + \frac{\partial \mathbf{x}}{\partial Z}\otimes \mathbf{k},\label{2_2}
\end{align}
\end{subequations}
where $\mathbf{\zeta}=R\mathbf{e}_R + \Theta \mathbf{e}_\Theta $ and $\mathbf{k}$ is the unit normal to the surface $\Omega$ in $\mathcal{B}_0$. 
The deformation gradient tensor can be decomposed as \citep{rodriguez1994stress}
\begin{align}
\mathbf{F}=\mathbf{AG},\label{decomp_F}
\end{align}  
where $\mathbf{G}$ represents growth tensor and $\mathbf{A}$ represents the elastic deformation tensor that ensures compatibility.
We also assume the plate to be incompressible and the incompressibility constraint is given by
\begin{align}
L({\mathbf{F,G}})=L_0(\mathbf{FG}^{-1})=L_0(\mathbf{A})=\text{det}(\mathbf{A})-1=0. \label{incompressibility_cond}
\end{align}
The energy density per unit volume $\phi$ of the material is
\begin{align}
\phi(\mathbf{F}, \mathbf{G}) = J_G\phi_0(\mathbf{FG}^{-1}),\label{energy_density}
\end{align}
where $J_G = \det(\mathbf{G})=\det(\mathbf{F})$ describes the local change in volume due to growth and $\phi_0 (\mathbf{FG}^{-1})$ is the elastic strain energy density. The external work done by the traction is given as
\begin{align}
V=\int_{\Omega}\mathbf{q}^{-} \cdot \mathbf{x}({\zeta},0)dA+\int_{\Omega}\mathbf{q}^{+} \cdot \mathbf{x}({\zeta},2h)dA+\int_{\partial \Omega_q}\int_{0}^{2h}{\mathbf{\tilde{q}}}\cdot\mathbf{x}(s,Z)dsdZ, \label{ext_work_done}
\end{align}
where $\mathbf{q}^{+}(\text{respectively} ~\mathbf{q}^-)$ represents the applied traction on top (respectively bottom) surface of region and $\mathbf{\tilde{q}}$ represents the traction on the lateral surface $\partial \Omega_q \times [0,2h]$, $\partial \Omega_q$ being the
boundary along the lateral surface,
and the symbol ($\cdot$) denotes the inner product.
If we neglect the body force, the total potential energy functional ($\psi$) for incompressible plate structure is
\begin{align}
\psi[\mathbf{x}(\mathbf{X}),p(\mathbf{X})]=\int_{\Omega}\int_{0}^{2h}J_G\phi_0(\mathbf{FG}^{-1}) dV -\int_{\Omega}\int_{0}^{2h} \big[ J_G~ p(\mathbf{X})L_0(\mathbf{FG}^{-1})\big]dV - V ,\label{energy_functional}
\end{align}
where $p(\mathbf{X})$ is the Lagrange multiplier, associated with the incompressibility constraint. 
We apply the principle of minimum potential energy and vanishing of the first variation with respect to $\mathbf{x}$ and $p$ of the above functional results in the governing equations
\begin{subequations}
\begin{align}
&\text{Div}\mathbf{P}=\mathbf{0}, \hspace{2.74in} \text{in}~\Omega \times  [0,2h], \label{gov_eq_a}\\
&\mathbf{Pk}\big|_{Z=0}=-\mathbf{q}^{-}(\mathbf{\zeta}),\hspace{0.6in}\left.\mathbf{Pk}\right|_{Z=2h}=\mathbf{q}^{+}(\zeta), \hspace{0.5in} \text{on}~\Omega, \label{gov_eq_b}\\
&\mathbf{Pn}=\mathbf{\tilde{q}}(s,Z),\hspace*{2.55in} \text{on}~ \partial  \Omega_q \times [0,2h].\label{gov_eq_c}
\end{align}
\end{subequations}
and the incompressibility constraint \eqref{incompressibility_cond}.
Here, $\mathbf{n}$ is the unit outward normal to the boundary $\partial \Omega_q$ and $\mathbf{P}=J_G \big[ \displaystyle\frac{\partial \phi_0}{\partial \mathbf{A}} $ -$p \frac{\partial L_0}{\partial \mathbf{A}}\big]\mathbf{G}^{-T}$ is recognised as the first Piola Kirchhoff stress tensor.
While deriving these equations, we have used the assumption that the rate of deformation of the growth process is very small compared to the elastic deformation \changes{\citep{amar2005growth, wang2018consistent}}, therefore the growth tensor $\mathbf{G}$ \changesg{is assumed to be constant in time}. Auxiliary calculations are presented in \hyperref[Appendix_C] { Appendix \ref{Appendix_C}}.


\subsection{Specialisation to two dimensions} \label{section_3}
To obtain the 2-D formulation of circular plate, we perform the series expansion of $\mathbf{x} ~\text{and}~ p$ in terms of $Z$ about the bottom surface, $Z=0$ following the approach by \cite{wang2018consistent, wang2019shape}
\begin{align}
\mathbf{x}(\mathbf{X}) 
&=\mathbf{x}^{(0)}(\mathbf{\mathbf{\zeta}})+Z\mathbf{x}^{(1)}(\mathbf{\mathbf{\zeta}})+\frac{Z^2}{2} \mathbf{x}^{(2)}(\mathbf{\mathbf{\zeta}})+\frac{Z^3}{3!}\mathbf{x}^{(3)}(\mathbf{\mathbf{\zeta}})+\frac{Z^4}{4!}\mathbf{x}^{(4)}(\mathbf{\mathbf{\zeta}})+O(Z^5), \label{3_1}\\
p(\mathbf{X})&=p^{(0)}(\mathbf{\zeta})+Zp^{(1)}(\mathbf{\zeta})+\frac{Z^2}{2} p^{(2)}(\mathbf{\zeta})+\frac{Z^3}{3!}p^{(3)}(\mathbf{\zeta})+\frac{Z^4}{4!}p^{(4)}(\mathbf{\zeta})+O(Z^5),\label{3_2}
\end{align}
where $(\cdot)^{(n)}=\displaystyle\left.\frac{\partial ^n {(\cdot)}}{\partial Z^n}\right|_{Z=0}~(n=0,1,2,3,4)$. Likewise, we write the Taylor's expansion of the deformation gradient $\mathbf{F}$, the elastic deformation tensor $\mathbf{A}$ and the inverse transpose of the growth tensor $\mathbf{G}$ as
\begin{subequations}
\begin{align}
\mathbf{F}&=\mathbf{F}^{(0)}(\zeta)+Z \mathbf{F}^{(1)}(\zeta)+\frac{Z^2}{2} \mathbf{F}^{(2)}(\zeta)+\frac{Z^3}{3!}\mathbf{F}^{(3)}(\zeta)+O(Z^4),\label{Taylor_F}\\
\mathbf{A}&=\mathbf{A}^{(0)}(\zeta)+Z\mathbf{A}^{(1)}(\zeta)+\frac{Z^2}{2} \mathbf{A}^{(2)}(\zeta)+\frac{Z^3}{3!}\mathbf{A}^{(3)}(\zeta)+O(Z^4),\label{Taylor_A}\\
\mathbf{G}^{-T}&=\bar{\mathbf{G}}^{(0)}(\zeta)+Z\bar{\mathbf{G}}^{(1)}(\zeta)+\frac{Z^2}{2} \bar{\mathbf{G}}^{(2)}(\zeta)+\frac{Z^3}{3!}\bar{\mathbf{G}}^{(3)}(\zeta)+O(Z^4).\label{Taylor_G}
\end{align}
\end{subequations}
The recursion relation for the expansion coefficients in \eqref{Taylor_F} is expressed as
\begin{align}
\mathbf{F}^{(n)}=\nabla{\mathbf{x}^{(n)}}+\mathbf{x}^{(n+1)}\otimes \mathbf{k}.~~~({n=0,1,2,3}). \label{Recursion_F}
\end{align}
Using \eqref{decomp_F}
and \eqref{Taylor_F} - \eqref{Taylor_G} we obtain the expressions for higher derivatives of $\mathbf{A}$ as
\begin{subequations}
\begin{align}
\mathbf{A}^{(0)}&=\mathbf{F}^{(0)}\bar{\mathbf{G}}^{{(0)}^{T}},\\
\mathbf{A}^{(1)}&=\mathbf{F}^{(0)}\bar{\mathbf{G}}^{{(1)}^{T}}+\mathbf{F}^{(1)}\bar{\mathbf{G}}^{{(0)}^{T}},\\
\mathbf{A}^{(2)}&=\mathbf{F}^{(0)}\bar{\mathbf{G}}^{{(2)}^{T}}+2\mathbf{F}^{(1)}\bar{\mathbf{G}}^{{(1)}^{T}}+\mathbf{F}^{(2)}\bar{\mathbf{G}}^{{(0)}^{T}}, \\
\mathbf{A}^{(3)}&=\mathbf{F}^{(0)}\bar{\mathbf{G}}^{{(3)}^{T}}+3\mathbf{F}^{(1)}\bar{\mathbf{G}}^{{(2)}^{T}}+3\mathbf{F}^{(2)}\bar{\mathbf{G}}^{{(1)}^{T}}+\mathbf{F}^{(3)}\bar{\mathbf{G}}^{{(0)}^{T}}.  
\end{align} 
\end{subequations}
Similarly, the expansion of the first Piola Kirchhoff stress tensor $\mathbf{P}$ is 
\begin{align}
\mathbf{P}(\mathbf{x},p)&=\mathbf{P}^{(0)}(\mathbf{x},p)+Z \mathbf{P}^{(1)}(\mathbf{x},p)+\frac{Z^2}{2!} \mathbf{P}^{(2)}(\mathbf{x},p)+\frac{Z^3}{3!}\mathbf{P}^{(3)}(\mathbf{x},p)+O(Z^4), \label{Taylor_P}
\end{align}
which can be written in component form ($[\mathbf{P}]_{ij} = P_{ij}$) as
\begin{align}
P_{ij}^{(0)}&=J_G\bigg[\mathcal{A}^{(0)}-p^{(0)}\mathcal{L}^{(0)}\bigg]_{i\alpha}\bar{G}^{(0)}_{\alpha j}, \label{Piola_0}\\
P_{ij}^{(1)}&=J_G \left[\bigg[\left[\mathcal{A}^{(1)}-p^{(0)}\mathcal{L}^{(1)}\right]_{i \alpha k \beta} A^{(1)}_{k \beta }-p^{(1)} \mathcal{L}^{(0)}_{i \alpha}\bigg]\bar{G}^{(0)}_{\alpha j} \changes{+}\left[\mathcal{A}^{(0)}-p^{(0)}\mathcal{L}^{(0)}\right]_{i\alpha}\bar{G}^{(1)}_{\alpha j}\right],\label{Piola_1}\\
P^{(2)}_{ij}&=J_{G}\bigg[\bigg[\mathcal{A}^{(1)}_{i k \alpha \beta}A^{(2)}_{\alpha \beta}+\big[\mathcal{A}^{(2)}-p^{(0)}\mathcal{L}^{(2)}\big]_{i k \alpha \beta m n}A^{(1)}_{\alpha \beta}A^{(1)}_{m n}-2p^{(1)}\mathcal{L}^{(1)}_{i k \alpha \beta}A^{(1)}_{\alpha \beta}-p^{(0)}\mathcal{L}^{(1)}_{i k \alpha \beta}A^{(2)}_{\alpha \beta}\nonumber \\
  & -p^{(2)}\mathcal{L}^{(0)}_{ik}\bigg]\bar{G}^{(0)}_{kj}+\bigg[2\mathcal{A}^{(1)}_{i k \alpha \beta}A^{(1)}_{\alpha \beta}-2 p^{(0)}\mathcal{L}^{(1)}_{i k \alpha \beta}A^{(1)}_{\alpha \beta}-2 p^{(1)}\mathcal{L}^{(0)}_{ik}\bigg] \bar{G}^{(1)}_{k j}\nonumber\\
  &+  \bigg[\mathcal{A}^{(0)}-p^{(0)}\mathcal{L}^{(0)}\bigg]_{i k}\bar{G}^{(2)}_{kj}\bigg],\label{Piola_2}
\end{align} 
with $\displaystyle \pmb{\mathcal{A}^{{i}}}(\mathbf{A}^{(0)})= \left.\frac{\partial^{i+1} \phi_0(\mathbf{A})}{\partial \mathbf{A}^{i+1}}\right|_{\mathbf{A}=\mathbf{A}^{(0)}}$ and~~ $\displaystyle \pmb{\mathcal{L}}^{{i}}(\mathbf{A}^{(0)})=\left.\frac{\partial L_0(\mathbf{A})}{\partial \mathbf{A}^{i+1}}\right|_{\mathbf{A}=\mathbf{A}^{(0)}}$. 
Further mathematical details of above calculations are provided in \hyperref[Appendix_B] {Appendix \ref{Appendix_B}}.

The two-dimensional governing system has $\mathbf{x}^{(\cdot)}$ and ${p}^{(\cdot)}$ as  unknown functions. In order to form a closed system of equations, we write the boundary conditions \eqref{gov_eq_b} at the bottom surface ($Z=0$)
\begin{align}
\left.\mathbf{P}^{(0)}\mathbf{k}\right|_{Z=0}=\mathbf{P}^{(0)}(\mathbf{A})\mathbf{k}=-\mathbf{q}^{-},\label{bottom_trac}
\end{align}
and at top surface ($Z=2h$) 
\begin{align}
\left.\mathbf{P}\mathbf{k}\right|_{Z=2h}=\mathbf{P}^{(0)}\mathbf{k}+2h\mathbf{P}^{(1)}\mathbf{k}+2h^2\mathbf{P}^{(2)}\mathbf{k}+\frac{4}{3}h^3\mathbf{P}^{(3)}\mathbf{k}+O(h^4)=\mathbf{q}^{+}. \label{top_trac}
\end{align}
The stress equilibrium equation neglecting the body force and external traction is given by \eqref{gov_eq_a}
\begin{equation}
\begin{aligned}
&\text{Div} \mathbf{P}=\mathbf{0}, \\
&\nabla \cdot {\mathbf{P}}+\frac{\partial }{\partial Z}[{\mathbf{P}}\mathbf{k}]=\mathbf{0},
\end{aligned}
\end{equation}
where $\nabla$ is the two-dimensional differentiation operator. 
Upon use of \eqref{Taylor_P}, we obtain a recursion relation for the first Piola--Kirchhoff stress as
\begin{align}
\nabla\cdot {\mathbf{P}}^{(n)}+ {\mathbf{P}}^{(n+1)}\mathbf{k}=\mathbf{0}. \label{recurssion_stress}
\end{align}
The series expansion of unknown functions $\mathbf{x(X)}$ and $p(\mathbf{X})$ have 19 unknowns with $\mathbf{x}^{n} ~ (n=0,1,2,3,4)$ comprising 15 unknowns and $p^{n}~ (n=0,1,2,3)$ comprising 4 unknowns. Thus, a closed system of 19 equations for the solution of $\mathbf{x}$ and $p$ is derived from equilibrium equation, traction (bottom and top surface) boundary conditions (\ref{gov_eq_a} - \ref{gov_eq_c}) and incompressibility condition \eqref{incompressibility_cond} as
\begin{subequations}
\begin{align}
&L_0(\mathbf{A}^{(0)})=0,\\
&\pmb{\mathcal{L}}^{(0)}[\mathbf{A}^{(1)}]=0,\\
&\pmb{\mathcal{L}}^{(0)}[\mathbf{A}^{(2)}] + \pmb{\mathcal{L}}^{(1)}[\mathbf{A}^{(1)},~\mathbf{A}^{(1)}]=0,\\
& \pmb{\mathcal{L}}^{(0)}[\mathbf{A}^{(3)}]+3 \pmb{\mathcal{L}}^{(1)}[\mathbf{A}^{(1)},~\mathbf{A}^{(2)}]+ \pmb{\mathcal{L}}^{(2)}[\mathbf{A}^{(1)},~\mathbf{A}^{(1)},~\mathbf{A}^{(1)}]=0,
\end{align}
\end{subequations}
where $\pmb{\mathcal{L}}^{(0)}[\mathbf{A}^{(1)}]=\pmb{\mathcal{L}}^{(0)}:\mathbf{A}^{(1)}=\displaystyle \text{det}(\A)\A^{-T}:\A^{(1)}$.
On subtracting the top \eqref{top_trac} and bottom traction \eqref{bottom_trac} condition we obtain the equilibrium equation
\begin{align}
\nabla \cdot \bar{\mathbf{P}}=-\bar{\mathbf{q}},\label{eql_eq_Piola}
\end{align}
where
\begin{align*}
\bar{\mathbf{P}}&=\frac{1}{2h}\int_{0}^{2h}\mathbf{P}dZ=\mathbf{P}^{(0)}+h\mathbf{P}^{(1)}+\frac{2}{3}h^2\mathbf{P}^{(2)}+O(h^3),\\
\bar{\mathbf{q}}&=\frac{\mathbf{q}^+ + \mathbf{q}^{-}}{2h}.
\end{align*}

\noindent $\bar{\mathbf{P}}$ is the average stress over the thickness, $\bar{\mathbf{q}}$ is the effective body force due to traction at top and bottom surface (see \hyperref[Appendix_C] {Appendix \ref{Appendix_C}}). \changes{ Using Taylor's expansion, the equilibrium equation \eqref{eql_eq_Piola} can be rewritten as \citep{wang2019uniformly}
\begin{equation}
\left.
\begin{aligned}
\nabla \cdot \mathbf{P}^{(0)}_{t} + h \nabla \cdot \mathbf{P}^{(1)}_{t} +\frac{2}{3} h^2 \nabla \cdot \mathbf{P}^{(2)}_{t} + O(h^3)=-\bar{\mathbf{q}}_{t}, \\
\big[ \nabla \cdot \bar{\mbf{P}} \big] \cdot \mbf{k} = \nabla \cdot \left[{\mathbf{P}^{(0)}}^T \mbf{k} \right] + h \nabla \cdot \left[{\mathbf{P}^{(1)}}^T \mbf{k} \right] + \frac{2}{3} h^2 \nabla \cdot \left[{\mathbf{P}^{(2)}}^T \mbf{k} \right] + O(h^3)=-\bar{{q}}_3,
\end{aligned} \right\} \label{decom_plate_eq_1}
\end{equation}
where the subscript `$t$' represents the in-plane (or tangential) component of a vector or tensor, ${\mathbf{q}}_t = q_1 \mathbf{e}_R + q_2 \mathbf{e}_\Theta,~ \bar{\mathbf{q}}_t = \displaystyle  \frac{\mathbf{q}_t^+ + \mathbf{q}_t^-}{2h} $ and $\bar{q}_3 = \displaystyle \frac{q_3^+ +q_3^-}{2h} $. Physically, Eq. \eqref{decom_plate_eq_1} represents the balance of forces}. The explicit expressions for $\mathbf{x}^{(2)}$, $p^{(1)}~\text{and}~ \mathbf{x}^{(3)},~p^{(2)}$ in terms of $\mathbf{x}^{(0)},~\mathbf{x}^{(1)}~\text{and}~p^{(0)}$ are given as
\begin{align}
\mathbf{x}^{(2)}&=-\mathbf{B}^{-1}\mathbf{f}^{(2)}-D^{{(0)}^{-1}} \pmb{\mathcal{L}}^{(0)} \left[\nabla \mathbf{x}^{(1)}\bar{\mathbf{G}}^{{(0)}^{T}}-\mathbf{B}^{-1}\mathbf{f}^{(2)}\otimes \bar{\mathbf{G}}^{(0)} \mathbf{k}+\mathbf{F}^{(0)}\bar{\mathbf{G}}^{{(1)}^{T}}\right]\mathbf{B}^{-1} \pmb{\mathcal{L}}^{(0)} \widehat{\mathbf{G}}^{(0)} \mathbf{k},\\
p^{(1)} & = -D^{{(0)}^{-1}} \pmb{\mathcal{L}}^{(0)} \left[\nabla \mathbf{x}^{(1)}\bar{\mathbf{G}}^{{(0)}^{T}}-\mathbf{B}^{-1}\mathbf{f}^{(2)}\otimes \bar{\mathbf{G}}^{(0)} \mathbf{k} +\mathbf{F}^{(0)}\bar{\mathbf{G}}^{{(1)}^{T}}\right],\\
\mathbf{x}^{(3)} & = \changes {-\mathbf{B}^{-1} \mathbf{f}^{(3)} + p^{(2)} \mathbf{B}^{-1} \pmb{\mathcal{L}}^{(0)} \widehat{\mathbf{G}}^{(0)} \mathbf{k} + 2p^{(1)} \mathbf{B}^{-1} \pmb{\mathcal{L}}^{(1)}[ \mathbf{A}^{(1)}] \widehat{\mathbf{G}}^{(0)} \mathbf{k} + 2 p^{(1)} \mbf{B}^{-1} \pmb{\mathcal{L}}^{(0)} \widehat{\mathbf{G}}^{(1)} \mathbf{k}},\\
p^{(2)}&=-D^{{(0)}^{-1}}\bigg[\pmb{\mathcal{L}}^{(1)}[\mathbf{A}^{(1)},\mathbf{A}^{(1)}]+ \pmb{\mathcal{L}}^{(0)}\left[-\mathbf{B}^{-1}\mathbf{f}^{(3)}\otimes  \bar{\mathbf{G}}^{(0)}\mathbf{k}\right] \nonumber \\ 
& \quad +2p^{(1)} \pmb{\mathcal{L}}^{(0)}\left[\mathbf{B}^{-1}\pmb{\mathcal{L}}^{(1)}[\mathbf{A}^{(1)}] \widehat{\mathbf{G}}^{{(0)}}\mathbf{k}\otimes\bar{\mathbf{G}}^{{(0)}}\mathbf{k}\right]
 +2p^{(1)} \pmb{\mathcal{L}}^{(0)}\left[\mathbf{B}^{-1} \pmb{\mathcal{L}}^{(0)}\widehat{\mathbf{G}}^{(1)} \mathbf{k}\otimes  \bar{\mathbf{G}}^{(0)}\mathbf{k}\right] \nonumber \\
 & \quad + \pmb{\mathcal{L}}^{(0)}\left[\nabla \mathbf{x}^{(2)}\bar{\mathbf{G}}^{{(0)}^{T}} +2\mathbf{F}^{(1)}\bar{\mathbf{G}}^{{(1)}^{T}}+\mathbf{F}^{(0)}\bar{\mathbf{G}}^{{(2)}^{T}}\right]\bigg],
\end{align}
where $ \widehat{\mathbf{G}}^{{(0)}} = J_G \bar{\mathbf{G}}^{{(0)}}$ and
\begin{align}
&{B}_{\alpha \beta}=J_G\left[\pmb{\mathcal{A}}^{(1)}-p^{(0)} \pmb{\mathcal{L}}^{(1)}\right]_{\alpha i \beta j}\left[\bar{\mathbf{G}}^{(0)}\mathbf{k}\right]_{i} \left[\bar{\mathbf{G}}^{(0)}\mathbf{k}\right]_{j}, \\
&{D}^{(0)}= \pmb{\mathcal{L}}^{(0)}\left[J_G \mathbf{B}^{-1} \pmb{\mathcal{L}}^{(0)} \bar{\mathbf{G}}^{(0)}\mathbf{k} \otimes \bar{\mathbf{G}}^{(0)}\mathbf{k}\right],\\
&\mathbf{f}^{(2)}=\nabla~.~ \mathbf{P}^{(0)}+\bigg[\left[ \pmb{\mathcal{A}}^{(1)}-p^{(0)}\pmb{\mathcal{L}}^{(1)}\right]\left[\nabla \mathbf{x}^{(1)}\bar{\mathbf{G}}^{{(0)}^{T}}+\mathbf{F}^{(0)}\bar{\mathbf{G}}^{{(1)}^{T}}\right]\bigg] \widehat{\mathbf{G}}^{(0)}\mathbf{k} 
+\left[ \pmb{\mathcal{A}}^{(0)} - p^{(0)} \pmb{\mathcal{L}}^{(0)}\right]\widehat{\mathbf{G}}^{(1)}\mathbf{k}, \\
&\mathbf{f}^{(3)}=\nabla~.~ \mathbf{P}^{(1)}+
\bigg[\left[\pmb{\mathcal{A}}^{(1)}-p^{(0)} \pmb{\mathcal{L}}^{(1)}\right]\left[\nabla \mathbf{x}^{(2)}\bar{\mathbf{G}}^{{(0)}^{T}}+2 \mathbf{F}^{(1)} \bar{\mathbf{G}}^{{(1)}^{T}}+\mathbf{F}^{(0)} \bar{\mathbf{G}}^{{(2)}^{T}}\right]\bigg]  \widehat{\mathbf{G}}^{(0)} \mathbf{k} \nonumber\\
& \hspace{0.7in}+\bigg[\left[ \pmb{\mathcal{A}}^{(2)}-p^{(0)} \pmb{\mathcal{L}}^{(2)}\right]\left[ \mathbf{A}^{(1)},\mathbf{A}^{(1)}\right]\bigg] \widehat{\mathbf{G}}^{(0)} \mathbf{k}
+\bigg[\left[ \pmb{\mathcal{A}}^{(1)}-p^{(0)}\pmb{\mathcal{L}}^{(1)}\right]\left[\mathbf{A}^{(1)}\right]\bigg] \widehat{\mathbf{G}}^{(1)}\mathbf{k}
\nonumber \\
&\hspace{3.25in} +\left[\pmb{\mathcal{A}}^{(0)}-p^{(0)} \pmb{\mathcal{L}}^{(0)}\right]  \widehat{\mathbf{G}}^{(2)}\mathbf{k}. 
\end{align}

\section{Growth induced deformation} \label{section_4}
In this section, we discuss two cases of circular hyperelastic plate growing in only radial direction \changes{\citep{wu2015modelling}} and \changes{combined} radial and circumferential directions. We assume a constant growth function in both the cases.

\subsection{ {Radial growth }}
Consider a thin isotropic circular plate which undergoes axisymmetric deformation ($\theta=\Theta$) with the position vector $\mathbf{x}$($r,\theta,z$) in deformed configuration. The series expansion of unknown functions $ r(R,Z)$, $z(R,Z) ~\text{and}~p(R,Z)$ in terms of $Z$ is written as
\begin{align}
r(R,Z)&=r^{(0)}(R)+Zr^{(1)}(R)+\frac{1}{2!}Z^2r^{(2)}(R)+\frac{1}{3!}Z^3r^{(3)}(R)+\frac{1}{4!}Z^4r^{(4)}(R)+O(Z^5),\\
z(R,Z)&=z^{(0)}(R)+Zz^{(1)}(R)+\frac{1}{2!}Z^2z^{(2)}(R)+\frac{1}{3!}Z^3z^{(3)}(R)+\frac{1}{4!}Z^4z^{(4)}(R)+O(Z^5),\\
p(R,Z)&=p^{(0)}+Zp^{(1)}(R)+\frac{1}{2!}Z^2p^{(2)}(R)+\frac{1}{3!}Z^3p^{(3)}(R)+\frac{1}{4!}Z^4p^{(4)}(R)+O(Z^5),
\end{align}
where we have used the notation $r^{(n)}=\displaystyle\frac{\partial^n r}{\partial Z^n}$, $z^{(n)}=\displaystyle\frac{\partial^n z}{\partial Z^n}$ and $p^{(n)}=\displaystyle\frac{\partial^n p}{\partial Z^n}$.

We assume the plate to follow neo-Hookean elastic constitutive law given by
\begin{align}
\phi_0(\A)=C_0\left[ \text{tr}(\mathbf{A}^T\mathbf{A})-3\right], \label{neo_hook_energy}
\end{align}
where $C_0$ is the ground state shear modulus.
On substitution of \eqref{decomp_F} in \eqref{neo_hook_energy}, the first Piola Kirchhoff stress $\mathbf{P}$ 
is obtained as (see \hyperref[Appendix_D]{Appendix \ref{Appendix_D} })
\begin{align}
\mathbf{P}=J_G \bigg[2C_0\mathbf{A}-p \mathbf{A}^{-T}\bigg]\mathbf{G}^{-T}. \label{4_6}
\end{align}
\changes{We can rewrite \eqref{decom_plate_eq_1} in component form as}
\changes{
\begin{align}
& P_{rR,R}^{(0)} + \frac{1}{R} P_{r\Theta,\Theta}^{(0)} + \frac{1}{R} \left[ P_{rR}^{(0)} \right] + h \bigg[P_{rR,R}^{(1)} + \frac{1}{R} P_{r\Theta,\Theta}^{(1)} + \frac{1}{R} \left[ P_{rR}^{(1)} \right]\bigg] 
 +O(h^2)=-\bar{{q}}_{1}, \\
&P_{\theta R,R}^{(0)} + \frac{1}{R} P_{\theta\Theta,\Theta}^{(0)} + \frac{1}{R} \left[ P_{\theta R}^{(0)} \right] + h \bigg[P_{\theta R,R}^{(1)} + \frac{1}{R} P_{\theta\Theta,\Theta}^{(1)} + \frac{1}{R} \left[ P_{\theta R}^{(1)} \right]\bigg]  +O(h^2)=-\bar{{q}}_{2}, \\
&P_{z R,R}^{(0)} + \frac{1}{R} P_{z \theta,\Theta}^{(0)} + \frac{1}{R} \left[ P_{z R}^{(0)} \right] + h \bigg[P_{z R,R}^{(1)} + \frac{1}{R} P_{z\Theta,\Theta}^{(1)} + \frac{1}{R} \left[ P_{z R}^{(1)} \right]\bigg]  +O(h^2)=-\bar{{q}}_{3}. 
\end{align} }
We can substitute the expression of stress from \eqref{eql_eq_Piola}  and omit the higher order terms in $h$ to get
\begin{align}
2C_0 \nabla \cdot \left[J_G \mathbf{F}^{(0)}\bar{\mathbf{G}}^{{(0)}^{T}}\bar{\mathbf{G}}^{(0)}\right]-\nabla \cdot \left[p^{(0)} J_G \mathbf{F}^{{(0)}^{-T}}\bar{\mathbf{G}}^{(0)}\right]+ O(h)=-\bar{\mathbf{q}}. \label{expand_avg_Piola}
\end{align}

Consider constant radial growth given by the
growth tensor $ \mathbf{G}=\text{diag}(\lambda,1,1)$
that results in
\begin{align}
\big[ \bar{\mathbf{G}}^{(0)} \big]=\begin{bmatrix}
\displaystyle \frac{1}{\lambda} & 0 & 0\\
0 & 1 & 0\\
0 & 0 & 1
\end{bmatrix}, 
\quad \big[  \mathbf{F}^{(0)} \big] = 
\begin{bmatrix}
{r^{(0)}}' & 0 & r^{(1)}\\
0 & \displaystyle\frac{r^{(0)}}{R} & 0\\
{z^{(0)}}' & 0 & z^{(1)}
\end{bmatrix} , \quad 
J_G=\text{det}({\mathbf{G}})=\lambda. \label{exp_F_G}
\end{align}
where a superposed prime  denotes  partial derivative with respect to $R$. 
Substituting \eqref{exp_F_G} in \eqref{expand_avg_Piola} we obtain the governing equation
\begin{align}
2C_0\nabla \cdot \lambda \begin{bmatrix}
{r^{(0)}}' & 0 & r^{(1)}\\
0 & \displaystyle\frac{r^{(0)}}{R} & 0\\
{z^{(0)}}' & 0 & z^{(1)}
\end{bmatrix} \begin{bmatrix}
\displaystyle \frac{1}{\lambda^2} & 0 & 0\\
0 & 1 & 0\\
0 & 0 & 1 \end{bmatrix}~-~ \nabla \cdot p^{(0)} \lambda \frac{\text{cofac}(\mathbf{F}^{(0)})}{\det \mathbf{F}^{(0)}}
+ O(h)=-\bar{\mathbf{q}}. \label{simplified_expand_avg_Piola}
\end{align}
Note that for the sake of brevity, we do not explicitly \changesg{represent} the \changes{$O(h)$ and $O(h^2)$} terms here but they are utilized for calculations later in \hyperref[sec: stability analysis radial] {Section \ref{sec: stability analysis radial}}.

\subsubsection{Plate supported by the Winkler foundation: Traction condition}

\begin{figure}
\centering
\includegraphics[width=0.7\linewidth]{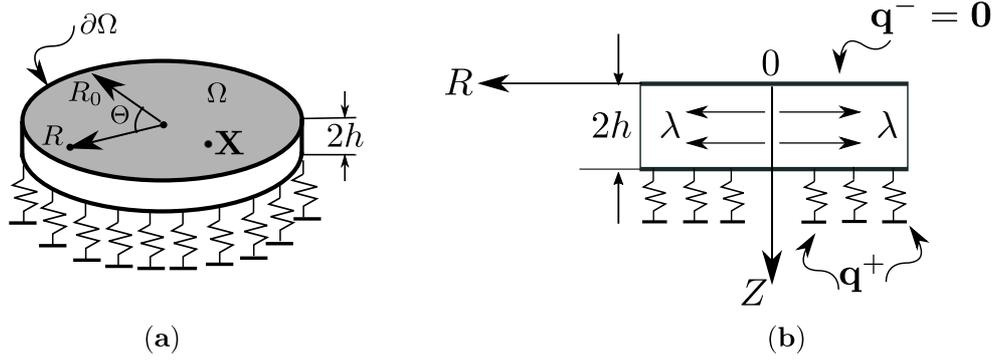}
\caption{Circular plate in reference configuration growing with growth factor $\lambda$ resting on a Winkler foundation. (a) Perspective view  (b) Front view.}\label{fig2}
\end{figure}

\changes{We concern ourselves with a circular plate resting on a Winkler foundation as shown in \hyperref[fig2] {Figure \ref{fig2}}. Winkler foundation models the elastic support provided to the growing plate}. The top surface ($Z=0$) of the plate is assumed to be traction-free and the bottom surface, is supported by the Winkler foundation which provides a {transverse load ${q}_3^{+}=-K_0 \lambda W_0$}, where $K_0$ is the elastic constant of the foundation and $W_0$ is the transverse ($Z$) component of the displacement. \changesg{$\lambda$ is the growth multiplier which }represents the fact that the traction 
is applied in the current grown configuration, as mass and stress state of the plate changes due to growth.

For this case, the components of the effective body force $(\bar{\mathbf{q}}$) are given by
\begin{align}
\bar{q}_{1}&=0, \\
\bar{q}_{3} 
&=-\frac{1}{2h}K_0 \lambda \bigg[z^{(0)} + 2hz^{(1)} +\frac{1}{2}~ [2h]^2 z^{(2)} + \frac{1}{6}~ [2h]^3 z^{(3)} - 2h \bigg]. \label{trac_Z}
\end{align}
The governing equations are obtained by substituting \eqref{trac_Z} in \eqref{simplified_expand_avg_Piola} 
\changes{
\begin{align}
&\frac{\partial }{\partial R}\left[2C_0 \frac{{r^{(0)}}'}{\lambda} -p^{(0)} \frac{r^{(0)} z^{(1)}}{R} \right] + \frac{1}{R} \bigg[ P_{rR}^{(0)} - p^{(0)} \left[ \frac{r^{(0)} z^{(1)}}{R} \right] \bigg]+O(h) =0 , \label{diff_gov_eq1}\\
&\frac{\partial }{\partial R} \bigg[2 C_0 \frac{{z^{(0)}}'}{\lambda}  - p^{(0)} \left[\frac{-r^{(0)} r^{(1)}}{R}\right] \bigg] + \frac{1}{R} \bigg[P_{zR}^{(0)} - p^{(0)} \left[ \frac{-r^{(0)} r^{(1)}}{R} \right] \bigg]
  + O(h) \nonumber\\ 
& \hspace{2in}= -K_0 \lambda \bigg[\frac{z^{(0)}}{2h} + z^{(1)} + h z^{(2)} + \frac{2}{3} h^2 z^{(3)} -1 \bigg]. \label{diff_gov_eq2}
\end{align} 
}
Assuming simply supported condition along the edge of circular plate, the boundary conditions at the center $R=0$ and the edge $R=R_0$ are given as 
\begin{equation}
r^{(0)}(0)=0, \quad r^{(0)}(R_0) =R_0, \quad
r^{(0)}(0) + [2h] r^{(1)}(0)  + \frac{1}{2} [2h^2] r^{(2)}(0) =0, \label{eqn: bc1 radial}
\end{equation}
where the unknown variables in terms of $r^{(0)}$ and $z^{(0)}$ are given as
\begin{align}
r^{(1)}= \displaystyle -\frac{\lambda {z^{(0)}}'}{\displaystyle \frac{r^{(0)}}{R}\left[ {z^{(0)}}'^2+{r^{(0)}}'^2\right]}, \quad
z^{(1)}&=\frac{\lambda {r^{(0)}}'}{\displaystyle  \frac{r^{(0)}}{R}\left[{z^{(0)}}'^2+ {r^{(0)}}'^2 \right]},\quad
p^{(0)}=\frac{2C_0\lambda^2}{{\left[ \displaystyle  \frac{r^{(0)}}{R} \right]^2\left[{z^{(0)}}'^2 + {r^{(0)}}'^2\right]}}. \label{exp_of_unknown_var}
\end{align}
The explicit expressions for $r^{(1)}, ~z^{(1)},~p^{(0)}$ are derived in \hyperref[Appendix_D] {Appendix \ref{Appendix_D}}.

\subsubsection{Stability analysis}
\label{sec: stability analysis radial}
The principal solution for plate deformation 
is given by
\begin{align}
r^{(0)}(R)=R, \quad z^{(0)}(R) = -2h[\lambda-1], \label{principle_sol}
\end{align}
where the second equation ensures that the $Z$ displacement of the lower surface vanishes, that is $z^{(0)} + 2hz^{(1)}-2h=0$ and in the homogeneous deformation we get the non zero quantity $z^{(1)}=\lambda$ (because $ {r^{(0)}}'=1 ~\text{and} ~{z^{(0)}}'=0$). 
A small perturbation in the homogeneous equilibrium state by a parameter $\epsilon$ results in
\begin{align}
r^{(0)}(R)=R+\epsilon \Delta U(R) \quad \text{and} \quad z^{(0)}(R)=-2h[\lambda-1]+\epsilon \Delta W(R). \label{ansatz}
\end{align}
\changes{
We define
\begin{align}
\mathbf{m} = \frac{\mbf{q}^+ - \mbf{q}^{-}}{2} = \mathbf{P}^{(0)} \mathbf{k} + h \mathbf{P}^{(1)} \mathbf{k} + h^2 \mathbf{P}^{(2)} \mathbf{k} +O(h^3). \label{trac_balance}
\end{align}
To simplify our calculations, we eliminate the terms with $\mbf{P}^{(2)}$ in equation \eqref{decom_plate_eq_1} by subtracting the divergence of \eqref{trac_balance} from \eqref{decom_plate_eq_1}$_2$  \citep{wang2019uniformly}. We keep the terms that correspond to bending and obtain
\begin{subequations}
\begin{align}
& \nabla \cdot \mathbf{P}^{(0)}_{t} + h\nabla \cdot \mathbf{P}^{(1)}_{t} =-\bar{\mathbf{q}}_{t}, \label{mod_gov_in_plane_plate_eq}\\
& \nabla \cdot \left[ ({\mbf{P}^{(0)}}^{T} \mbf{k}) - ({\mbf{P}^{(0)}} \mbf{k}) \right] + h \bigg[ \nabla \cdot \left[ ({\mbf{P}^{(1)}}^{T} \mbf{k}) - ({\mbf{P}^{(1)}} \mbf{k}) \right]\bigg] \nonumber\\
& \hspace{2in}+ \frac{1}{3} h^2 \nabla \cdot [\nabla \cdot {\mbf{P}_t}^{(1)}] = -\bar{q}_3 - \nabla \cdot \mbf{m}_t,\label{mod_gov_plate_eq}
\end{align}
\end{subequations}
where $\mbf{m}_t$ is the tangential component of  $\mbf{m}$. Physically, Eq. \eqref{trac_balance} represents the balance of moments.}

\noindent We define the dimensionless quantities 
\begin{align}
\rho=\frac{R}{R_0}, \hspace{0.1in} \bar{h}=\frac{h}{R_0} \changes{,} \hspace{0.1in} U=\frac{\Delta U}{R_0}, \hspace{0.1in} W=\frac{\Delta W}{R_0} , \label{non_dim_parm}
\end{align}
where $\rho \in [0,1]$ and $R_0$ is radius of circular plate in the reference configuration.
On substituting the ansatz \eqref{ansatz}  in \changes{ \eqref{diff_gov_eq1}$_1$ and \eqref{mod_gov_plate_eq}, simplifying using \eqref{exp_of_unknown_var}, \eqref{non_dim_parm}} and collecting only $O (\epsilon)$ terms
\changes{
\begin{align}
&\frac{2}{\lambda}[1+3\lambda^4] {U''}+\frac{2}{\rho \lambda} [1 + 5 \lambda^4] U' - \bar{h} \bigg[4[1 + \lambda^4] W''' 
 + \frac{2}{\rho} \big[ 1 + 5 \lambda^4 \big] \bigg]  = 0. \label{non_dim_diff_eq1}
 \end{align}
\begin{align}
& \frac{2}{\lambda}[1-\lambda^4] {W''} + \frac{2}{\rho \lambda} [1-\lambda^4] W' \nonumber \\
& - \bar{h} \bigg[ \frac{4}{\rho \lambda} \big[ 4 \lambda^5 - 4 \lambda^4 + \lambda - 1 \big] U'' + \frac{4}{\rho^2} \big[ \lambda^4 - 1 \big] U' - \frac{4}{\rho^3} \big[ \lambda^4 - 1 \big] U \bigg] \nonumber\\
& - \frac{2}{3} \bar{h}^2 \bigg[ 2 [1 + \lambda^4] W^{iv} + \frac{1}{\rho} \big[ 3 + 7 \lambda^4 \big] W''' \bigg]   \nonumber \\ 
& - \frac{\beta \lambda}{2 \bar{h}} {W} + \beta \lambda^2 U' + \frac{\beta \lambda^2}{\rho} U - \frac{ \bar{h} \beta [2 \lambda^4 - 1]}{\rho \lambda} W' - \bar{h} \beta \lambda^3  W'' \nonumber\\
& - \frac{2}{3} \bar{h}^2 \beta \bigg[\frac{4}{\rho} \big[ 1 + \lambda^4 \big] U'' - \frac{1}{\rho^2} \big[ 2 \lambda^4 - 3 \big] U' + \frac{1}{\rho^3} \big[ 2 \lambda^4 - 3 \big] U \bigg] = 0 .\label{non_dim_diff_eq3}
\end{align}
where $\beta=\displaystyle \frac{K_0}{C_0} R_0$ is a non-dimensional constant. The higher order derivative of in-plane displacement term is omitted for simplification. The plate boundary conditions \eqref{eqn: bc1 radial} is given as
\begin{equation}
\begin{aligned}
&U(0) = W'(0) = W'''(0) = 0,\\
&U(1) = W(1) = W''(1) = 0. \label{non_dim_bc}
\end{aligned}
\end{equation}
}
Equations \eqref{non_dim_diff_eq1} and \eqref{non_dim_diff_eq3} are coupled ODEs and can be written as a system of first order ODEs by defining 
\changes{
\begin{align}
 U=y_1, \qquad  U'=y_2, \qquad  W=y_3, \qquad  W'=y_4, \qquad  W''=y_5, \qquad  W'''=y_6. 
\end{align}  
}
The system of first order ordinary differential equations is written in the form of 
\begin{align}
\mathbf{Y}'=\pmb{\mathscr{A}}(\rho; \lambda, \bar{h}, \beta)\mathbf{Y}, \label{first_order_sys}
\end{align}
where $\mathbf{Y}=[\Delta U,~ \Delta U',~\Delta U'',~ \Delta W, ~\Delta W',~ \Delta W'']$~=~$[y_1 ~ y_2 ~ y_3~y_4 ~y_5 ~ y_6]$  and $\pmb{\mathscr{A}}$ is given by
\begin{align}
\pmb{\mathscr{A}}=\begin{bmatrix}
0 & 1 & 0 & 0 & 0 & 0\\
\mathscr{A}_{21} & \mathscr{A}_{22} &  \mathscr{A}_{23} &   \mathscr{A}_{24} & \mathscr{A}_{25} &   \mathscr{A}_{26}\\
 0 & 0 &  0 &   1 & 0 &   0\\
0 & 0 & 0 & 0 & 1 & 0\\
0 & 0 & 0 & 0 & 0 & 1 \\
  \mathscr{A}_{61} &  \mathscr{A}_{62} &   \mathscr{A}_{63} & \mathscr{A}_{64} & \mathscr{A}_{65} & \mathscr{A}_{66}
\end{bmatrix}, \label{first_order_sys_coeff}
\end{align}
\changes{
where 
\begin{align*}
&\mathscr{A}_{21} =  0, \quad  \mathscr{A}_{22}=\frac{\lambda}{2 (1 + 3 \lambda^4)} \bigg[- \frac{2}{\rho \lambda} [1 + 5 \lambda^4]\bigg], \quad \mathscr{A}_{23} = 0, \quad \mathscr{A}_{24} = 0,\nonumber\\
& \mathscr{A}_{25} = \frac{\lambda}{2 (1 + 5 \lambda^4)} \bigg[\frac{2 \bar{h}}{\rho} \big[1 + 5 \lambda^4 \big] \bigg], \qquad \mathscr{A}_{26} = \frac{\lambda}{2 (1 + 3 \lambda^4)} \bigg[4 \bar{h} [1 + \lambda^4]\bigg] \\
& \mathscr{A}_{61} = \frac{3}{4 \bar{h^2} [1 + \lambda^4]} \Bigg[ \frac{4 \bar{h}}{\rho^3} [\lambda^4 - 1] +  \frac{\beta \lambda^2}{\rho} - \frac{2}{3} \frac{\bar{h}^2 \beta}{\rho^3} [2 \lambda^4 - 3] \Bigg],\nonumber\\
& \mathscr{A}_{62} = \frac{3}{4 \bar{h^2} [1 + \lambda^4]} \Bigg[-\frac{4h}{\rho \lambda} \left[\frac{\lambda [4 \lambda^5 - 4 \lambda^4 + \lambda - 1]}{2 [1 + 3 \lambda^4]}\right] \bigg[- \frac{2}{\rho \lambda} [1 + 5 \lambda^4]\bigg] - \frac{4 h }{\rho^2} [\lambda^4 - 1]  \nonumber\\
& \qquad \qquad  + \beta \lambda^2 - \frac{2}{3} \bar{h}^2 \beta \bigg[ \frac{4 \lambda [1 + \lambda^4]}{2 \rho [1 + 3 \lambda^4]} \bigg[- \frac{2}{\rho \lambda} [1 + 5 \lambda^4]\bigg] \bigg] + \frac{2}{3} \bar{h}^2 \frac{\beta}{\rho^2} [2 \lambda^4 - 3] \Bigg],  \nonumber\\
 & \mathscr{A}_{63} = \frac{3}{4 \bar{h^2} [1 + \lambda^4]} \bigg[-\frac{\beta \lambda}{2 \bar{h}}\bigg], \qquad \mathscr{A}_{64} = \frac{3}{4 \bar{h^2} [1 + \lambda^4]} \bigg[\frac{2}{\rho \lambda}[1 - \lambda^4] - \frac{h \beta}{\rho \lambda} [2 \lambda^4 - 1] \bigg], \nonumber\\
& \mathscr{A}_{65} = \frac{3}{4 \bar{h^2} [1 + \lambda^4]} \Bigg[ \frac{2}{\lambda} [1 - \lambda^4] -\frac{4h}{\rho \lambda} \left[\frac{\lambda [4 \lambda^5 - 4 \lambda^4 + \lambda - 1]}{2 [1 + 3 \lambda^4]}\right] \bigg[\frac{2 \bar{h}}{\rho} [1 + 5 \lambda^4] \bigg] - h \beta \lambda^3 \nonumber\\
& \qquad \qquad \qquad \qquad \qquad \qquad \qquad \qquad   -\frac{2}{3} \bar{h}^2 \beta \bigg[ \frac{4 \lambda [1 +  \lambda^4]}{2 \rho [1 + 3 \lambda^4]} \bigg[\frac{2 \bar{h}}{\rho} [1 + 5 \lambda^4] \bigg]\bigg] \Bigg], \nonumber\\
& \mathscr{A}_{66} = \frac{3}{4 \bar{h^2} [1 + \lambda^4]} \Bigg[ -\frac{4h}{\rho \lambda} \left[\frac{\lambda [4 \lambda^5 - 4 \lambda^4 + \lambda - 1]}{2 [1 + 3 \lambda^4]}\right] \bigg[ 4 \bar{h} [1 + \lambda^4] \bigg] - \frac{2}{3} \frac{\bar{h}^2}{\rho} [3 + 7 \lambda^4] \nonumber\\
&\qquad \qquad \qquad \qquad \qquad \qquad \qquad \qquad  -\frac{2}{3} \bar{h}^2 \beta \bigg[ \frac{4 \lambda [1 +  \lambda^4]}{2 \rho [1 + 3 \lambda^4]}\bigg] \bigg[4 \bar{h} [1 + \lambda^4] \bigg] \Bigg].
\end{align*}
}
\noindent We first determine the critical value of growth factor ($\lambda_{cr}$) that results in the onset of a bifurcation and then we discuss the associated buckling modes. 
The system of first order ODEs \eqref{first_order_sys} is treated as two-point boundary value problem.
This stiff eigenvalue problem is solved using the 
compound matrix method \citep{ng1979numerical,ng1985compound,lindsay1992note, haughton1997eversion}.
Following the compound matrix approach, the system \eqref{first_order_sys} is converted into 20 first order equations of the form $\mathbf{\Phi}'=\pmb{\mathscr{A}}^{*} (\rho;\lambda, \bar{h}, \beta)\mathbf{\Phi}$ (A detailed description of the solution procedure is given in \hyperref[Appendix_A] {Appendix \ref{Appendix_A}})
\changes{
\begin{align}
\Phi_{1}'& =\Phi_2 + \mathscr{A}_{22} \Phi_1 - \mathscr{A}_{24} \Phi_5 - \mathscr{A}_{25} \Phi_6 - \mathscr{A}_{26} \Phi_{7}, \nonumber \\
\Phi_2'& = \Phi_3 + \mathscr{A}_{22} \Phi_2 + \mathscr{A}_{23} \Phi_5 - \mathscr{A}_{25} \Phi_8 - \mathscr{A}_{26} \Phi_{9}, \nonumber\\
\Phi_3'& =\Phi_4 + \mathscr{A}_{22} \Phi_3 + \mathscr{A}_{23} \Phi_6 + \mathscr{A}_{24} \Phi_8 - \mathscr{A}_{26} \Phi_{10}, \nonumber\\
\Phi_4'& = \mathscr{A}_{22} \Phi_4 + \mathscr{A}_{23} \Phi_7 + \mathscr{A}_{24} \Phi_9 + \mathscr{A}_{25} \Phi_{10} + \mathscr{A}_{63} \Phi_{1} + \mathscr{A}_{64} \Phi_2 + \mathscr{A}_{65} \Phi_3 + \mathscr{A}_{66} \Phi_{4}, \nonumber\\
\Phi_{5}'& = \Phi_{11} +  \Phi_{6},\nonumber \\
\Phi_6'& =\Phi_{12}  + \Phi_8 +\Phi_7, \nonumber\\
\Phi_{7}'& = \Phi_{13} + \Phi_9 -  \mathscr{A}_{62} \Phi_{1} + \mathscr{A}_{64} \Phi_5 + \mathscr{A}_{65} \Phi_6 + \mathscr{A}_{66} \Phi_7, \nonumber\\
\Phi_8'& = \Phi_{14} + \Phi_{9} , \nonumber \\
\Phi_9'& = \Phi_{15} + \Phi_{10} - \mathscr{A}_{62} \Phi_{2} - \mathscr{A}_{63} \Phi_{5} + \mathscr{A}_{65} \Phi_{8} + \mathscr{A}_{66} \Phi_{9}, \nonumber\\
\Phi_{10}'& = \Phi_{16} - \mathscr{A}_{62} \Phi_{3} - \mathscr{A}_{63} \Phi_{6} - \mathscr{A}_{64} \Phi_{8} + \mathscr{A}_{66} \Phi_{10}, \label{compound_eqs} \\
\Phi_{11}'& = \Phi_{12} + \mathscr{A}_{21} \Phi_{5} + \mathscr{A}_{22} \Phi_{11} + \mathscr{A}_{25}\Phi_{17} + \mathscr{A}_{26} \Phi_{18}, \nonumber\\
\Phi_{12}'& = \Phi_{14} + \Phi_{13} + \mathscr{A}_{21} \Phi_{6} + \mathscr{A}_{22} \Phi_{12} - \mathscr{A}_{24} \Phi_{17} + \mathscr{A}_{26} \Phi_{19} , \nonumber\\
\Phi_{13}'& = \Phi_{15} + \mathscr{A}_{21} \Phi_{7} + \mathscr{A}_{22} \Phi_{13} - \mathscr{A}_{24} \Phi_{18} - \mathscr{A}_{25} \Phi_{19} + \mathscr{A}_{61} \Phi_{1} + \mathscr{A}_{64} \Phi_{11}+ \mathscr{A}_{65} \Phi_{12} + \mathscr{A}_{66} \Phi_{13} ,\nonumber \\
\Phi_{14}'& = \Phi_{15} + \mathscr{A}_{21} \Phi_{8} + \mathscr{A}_{22} \Phi_{14} + \mathscr{A}_{23} \Phi_{17} + \mathscr{A}_{26} \Phi_{20},\nonumber\\
\Phi_{15}'& = \Phi_{16} + \mathscr{A}_{21} \Phi_{9} + \mathscr{A}_{22} \Phi_{15} + \mathscr{A}_{23} \Phi_{18} - \mathscr{A}_{25} \Phi_{20} +\mathscr{A}_{61} \Phi_{2} - \mathscr{A}_{63} \Phi_{11} + \mathscr{A}_{65} \Phi_{14} + \mathscr{A}_{66} \Phi_{15}, \nonumber\\
\Phi_{16}'&= \mathscr{A}_{21} \Phi_{10} + \mathscr{A}_{22} \Phi_{16} + \mathscr{A}_{23} \Phi_{19} + \mathscr{A}_{24} \Phi_{20}+ \mathscr{A}_{61} \Phi_{3} - \mathscr{A}_{63} \Phi_{12}  - \mathscr{A}_{64} \Phi_{14} +\mathscr{A}_{66} \Phi_{16}, \nonumber\\
\Phi_{17}'& = \Phi_{18}, \nonumber\\
\Phi_{18}'& =  \Phi_{19} + \mathscr{A}_{61} \Phi_{5} + \mathscr{A}_{62} \Phi_{11} + \mathscr{A}_{65} \Phi_{17} + \mathscr{A}_{66} \Phi_{18}, \nonumber\\
\Phi_{19}'& =  \Phi_{20} + \mathscr{A}_{61} \Phi_{6} + \mathscr{A}_{62} \Phi_{12} - \mathscr{A}_{64} \Phi_{17} + \mathscr{A}_{66} \Phi_{19}, \nonumber \\
\Phi_{20}'&= \mathscr{A}_{61} \Phi_{8} + \mathscr{A}_{62} \Phi_{14} + \mathscr{A}_{63} \Phi_{17} + \mathscr{A}_{66} \Phi_{20}. \nonumber
\end{align}
}
The initial condition for the system of equations \eqref{compound_eqs} is 
\begin{align}
\Phi(0)=\begin{bmatrix}
\Phi_{1}, \Phi_{2}, \Phi_{3}, \Phi_{4}, \Phi_{5}, \Phi_{6}, \Phi_{7}, \Phi_{8}, \Phi_{9}, \Phi_{10}, \Phi_{11}, \Phi_{12}, \Phi_{13}, \Phi_{14}, \Phi_{15}, \Phi_{16}, \Phi_{17},  \Phi_{18}, \Phi_{19}, \Phi_{20}
\end{bmatrix}. 
\end{align}
The target condition is achieved by having $\det (\mathbf{CM})=0$ in order to obtain the non-trivial solution, 
where matrix $\mathbf{C}$ corresponds to  the boundary condition at the edge of circular plate \eqref{non_dim_bc} and $\mathbf{M}$ is the solution matrix 
\begin{align}
\changes{
\mathbf{C}=\begin{bmatrix}
1 & 0 & 0 & 0 & 0 & 0\\
0 & 0 & 1 & 0 & 0 & 0\\
0 & 0 & 0 & 0 & 1 & 0
\end{bmatrix}}~~~~ \text{and}~~~~ \mathbf{M}= \begin{bmatrix}
y_1^{(1)} & y_1^{(2)} & y_1^{(3)} \\ 
y_2^{(1)} & y_2^{(2)} & y_2^{(3)} \\
y_3^{(1)} & y_3^{(2)} & y_3^{(3)} \\
y_4^{(1)} & y_4^{(2)} & y_4^{(3)} \\
y_5^{(1)} & y_5^{(2)} & y_5^{(3)} \\
y_6^{(1)} & y_6^{(2)} & y_6^{(3)}
\end{bmatrix}. \label{solution_matrix}
\end{align}

\noindent For the current case, the corresponding initial conditions using \eqref{non_dim_bc} are given \changes{by $\Phi(12)=1$} and rest all are zero. 
Then, we integrate the system numerically in the interval of $0 < \rho \leq 1 $ until we achieve the target condition on the other boundary which is given by \changes{$\det(\mathbf{CM})$=(1,3,5)=$\Phi(6)=0$ }(see \hyperref[Appendix_A] {Appendix \ref{Appendix_A}}). 
The main objective of this optimization problem is to determine the critical value of growth factor $\lambda_{cr}$ for which the target value \changes{$\Phi(6)$ is zero}. 

\subsubsection{Results and discussion}
In this section, the buckling behaviour of a circular hyperelastic plate due to radial growth rested on Winkler foundation is presented.
In order to validate our numerical scheme based on the compound matrix method, we first evaluate the onset of buckling of a rectangular plate under uni-directional growth for which an analytical solution has been provided by \cite{wang2018consistent}.

\subsubsection*{a) Comparison of numerical results and analytical results for rectangular plate}
\begin{figure}
\centering
\includegraphics[width=0.3\linewidth]{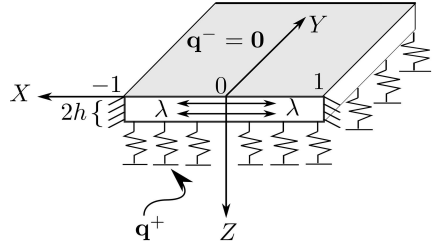}
\caption{Schematic of a rectangular plate resting on Winkler foundation and clamped at $X=\pm 1$.} \label{rec_fig}
\end{figure}

Consider a rectangular plate of thickness $2h$ clamped at the ends $X=\pm1$ and supported by a Winkler foundation as shown in \hyperref[rec_fig]{Figure \ref{rec_fig}}.
The plate grows along the X-axis with a growth stretch $\lambda$.
The compression effects due to the clamped boundary condition lead to buckling.

The governing plate differential equation is given by \citep{wang2018consistent}
\begin{align}
\psi_0 \Delta W + \psi_2 \Delta W'' + \psi_4 \Delta W''''=0, \label{rec_gov_eq}
\end{align}
where 
\begin{align*}
\psi_0&=\frac{\alpha}{2h},\\
\psi_2&=\frac{1}{\lambda^2 + 3\lambda^6} \bigg[ \lambda^4-1 \left[2+[6+h\alpha]\lambda^4\right] \bigg],\\
\psi_4&= \frac{4 h^2}{3+9\lambda^4} \bigg[3+h \alpha + [2+3 h \alpha] \lambda^4 + [3+2h\alpha] \lambda^8 \bigg],
\end{align*}
subjected to the boundary conditions
\begin{align}
\Delta W'(-1) = \Delta W'(1)=0 , ~~~~~~~ \Delta W'''(-1) = \Delta W'''(1)=0. \label{rec_bc_eq}
\end{align}

\begin{figure}
\centering
\begin{tabular}{c}
\includegraphics[width=0.7\linewidth]{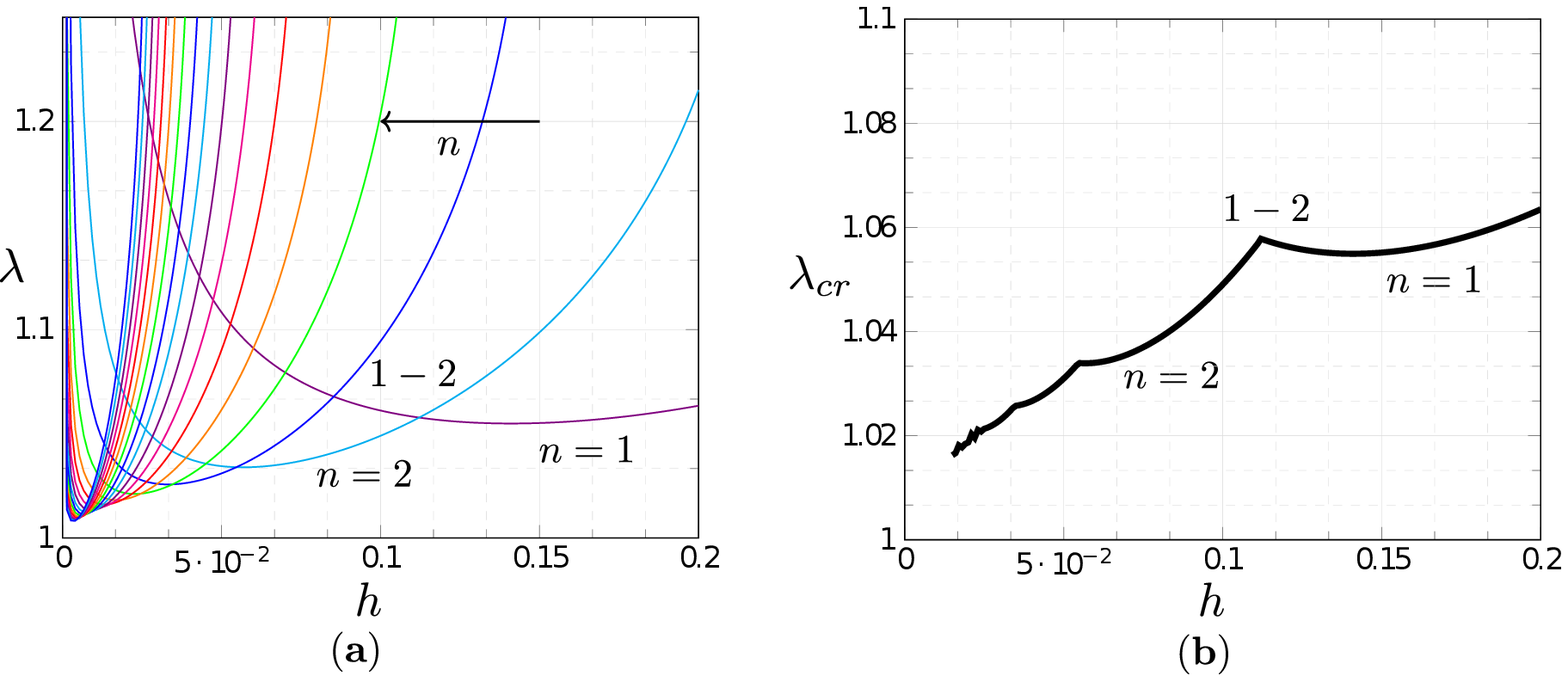} \\
\includegraphics[width=0.3\linewidth]{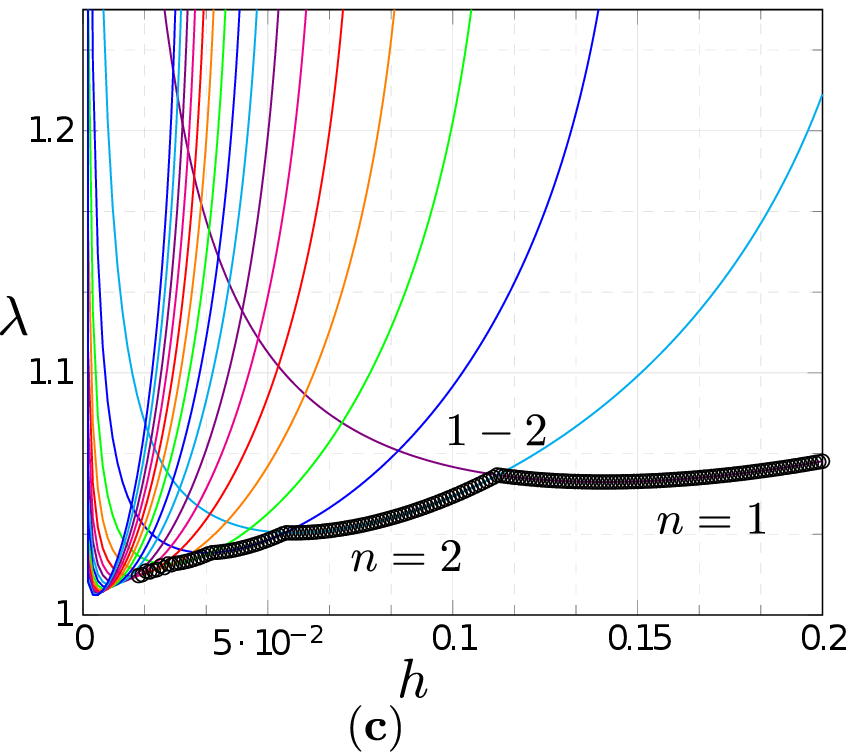}
\end{tabular} 
\caption{{ a) Bifurcation curves for various modes of rectangular plate \citep{wang2018consistent}. b) Variation of the \changesg{critical value of growth factor ($\lambda_{cr}$)} against thickness to length ratio ($h$) evaluated using the compound matrix method. c) Direct comparison of (a) and (b). }}\label{h and L variation for rec plate}
\end{figure}

The analytical results of bifurcation curves for various buckling modes is shown in \hyperref[h and L variation for rec plate]{Figure \ref{h and L variation for rec plate}a}. 
For thick plates ($h>0.15$), the first fundamental mode dominates. 
As we lower the value of $h$, the higher modes ($n=2,3, \dotsc $) are more stable. \hyperref[h and L variation for rec plate] {Figure \ref{h and L variation for rec plate}b} represents the variation of critical value of growth factor ($\lambda_{cr}$) with respect to thickness to length ratio $(h)$ \changes{obtained by solving \eqref{rec_gov_eq}--\eqref{rec_bc_eq} using compound matrix method.} \changes{Also, the numerical compound matrix method solution of \eqref{mod_gov_in_plane_plate_eq} and \eqref{mod_gov_plate_eq} specialized to a rectangular plate geometry provides the same result.} 
The $\lambda_{cr}$ variation is non-monotonous and has discontinuous derivatives at certain thickness values which suggests the phenomenon of mode jumping from high energy state to a lower one associated with changes in mode shape. 
We conclude that the peaks in 
\hyperref[h and L variation for rec plate] {Figure \ref{h and L variation for rec plate}b} that 
represent the transition of modes i.e., for thin plates, the higher modes are more stable than the lower modes. 
In \hyperref[h and L variation for rec plate]{Figure \ref{h and L variation for rec plate}c}, we superpose the numerical results on the analytical curves for direct comparison. 
It can be seen that the analytical and numerical results are in a near perfect agreement \changes{which shows the efficacy of the compound matrix method}.

\subsubsection*{b) Numerical results for circular plate with radial growth}

\begin{figure}
\centering
\includegraphics[width=0.7\textwidth]{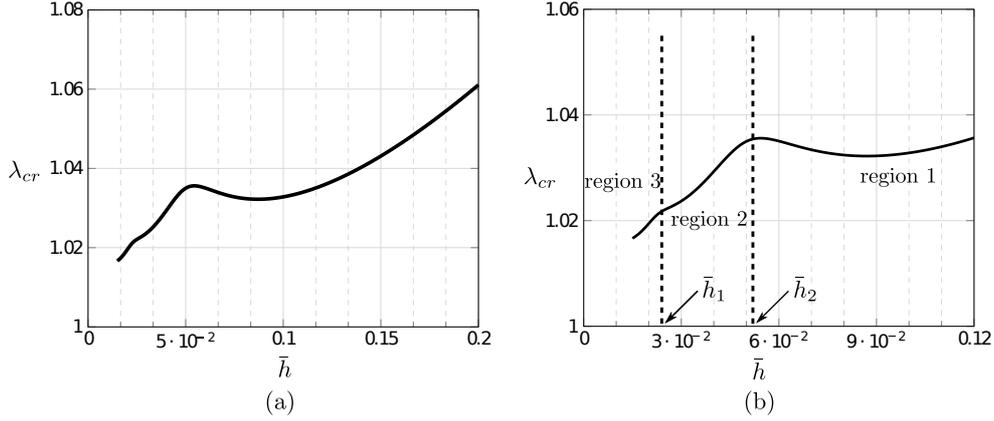}
\caption{\changes{{ a) Variation of critical growth factor $\lambda_{cr}$ with dimensionless thickness $\bar{h}$ for $\beta=0.2$. b) Enlarged part of (a). The thickness values $\bar{h}_1$ and $\bar{h}_2$ correspond to the critical points of mode switching.} }} \label{fig_circular_variation}
\end{figure}

The compound matrix method is used to compute the buckling parameter ($\lambda_{cr}$) of the circular hyperelastic plate under radial growth condition. 
The variation of critical growth factor ($\lambda_{cr}$) with respect to normalized thickness is shown in \hyperref[fig_circular_variation] {Figure \ref{fig_circular_variation}}. 
The results show the non-monotonous nature of $\lambda_{cr}$ variation and the points at which the mode jump phenomenon happens at certain plate thicknesses. 
Based on the above results, the graph can be divided into 3 regions of thickness in which a particular mode is dominant over the other modes. 
In the \changes{region 3}, which corresponds to lower plate thickness, a higher order mode and for higher plate thickness, the lower order mode dominates the buckling behaviour. Next, we determine the mode shapes corresponding to the 3 regions defined in \hyperref[fig_circular_variation] {Figure \ref{fig_circular_variation}}. 

To obtain the mode shapes, the Matlab ODE package \texttt{bvp4c} is used for solving the equations \eqref{non_dim_diff_eq1} and \eqref{non_dim_diff_eq3} subjected to boundary condition \eqref{non_dim_bc} at the critical point. 
Equations \eqref{non_dim_diff_eq1} and \eqref{non_dim_diff_eq3} are rewritten into a first order form of \changes{$\mathbf{Y}'
=\pmb{\mathscr{A}} \mathbf{Y}$, where  $\mathbf{Y}=[y1,~y2,~y3,~y4,~y5,~y6]=[ U,~  U',~ W, ~ W',~ W'',~W''']$} where $\pmb{\mathscr{A}}$  is given by \eqref{first_order_sys_coeff} and the boundary condition \eqref{non_dim_bc} is rewritten as
\changes{
\begin{align}
y_a(1)=0,~y_b(1)=0,~y_a(4)=0,~y_b(3)=0,~y_a(6)=0, ~y_b(5)=0. 
\end{align} 
}
where $y_a$ (respectively, $y_b$) define the  centre of the plate (respectively, edge of the plate).
\begin{figure}
\centering
\includegraphics[width=0.7\linewidth]{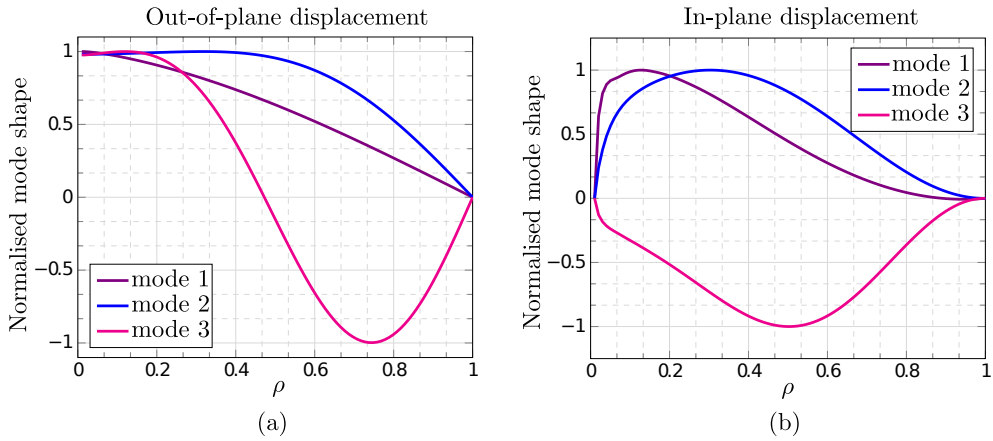}
\caption{ \changes{Normalised bifurcation mode shapes for circular plate under radial growth (a) {Normalised} out-of-plane displacement $W$. b) {Normalised} in-plane displacement $U$.}} \label{fig_mode_shapes}
\end{figure}

The mode shape results are obtained for a thickness value in the defined regions and the results are shown in \hyperref[fig_mode_shapes] {Figure \ref{fig_mode_shapes}}. In the region 3, a thickness \changes{$\bar{h} = 0.024$ is chosen for which $\lambda_{cr}=1.0218$ }is used to evaluate the mode shape of the circular hyperelastic plate. At this value of $\bar{h}$, a higher order mode shape (mode 3) exists and variation of out of plane displacement amplitude in radial direction is shown in \hyperref[fig_mode_shapes] {Figure \ref{fig_mode_shapes}}. Similarly, the modes in the region 2 and 1 represented by mode 2 and mode 1 are evaluated by choosing \changes{ $\bar{h} = 0.05 , \lambda_{cr} = 1.0350$ and $ \bar{h} =0.2 , \lambda_{cr}= 1.0611,$} respectively. 
Thus, we conclude that 
the buckling parameter ($\lambda_{cr}$) of circular plate under radial growth is non-monotonic with higher order modes in the thin plate regime and lower order modes in the thicker plate regime. 

\subsection{{Combined radial and circumferential growth }}
In this second case, \changes{ we assume a constant isotropic growth function $\lambda$ (i.e., $\lambda_{rr} = \lambda_{\theta \theta} = \lambda$) in both radial and circumferential direction \citep{wu2015modelling}. The isotropic growth tensor  $ \mathbf{G}=\text{diag}(\lambda,\lambda,1)$ results in the following  kinematics}

\begin{align}
\big[\bar{\mathbf{G}}^{(0)} \big]=
\begin{bmatrix}
\displaystyle \frac{1}{\lambda} & 0 & 0\\
0 & \displaystyle \frac{1}{\lambda} & 0\\
0 & 0 & 1
\end{bmatrix},\quad
 \big[\mathbf{F}^{(0)}\big]=
\begin{bmatrix}
\displaystyle\frac{\partial r^{(0)}}{\partial R} & \displaystyle \frac{1}{R}\frac{\partial r^{(0)}}{\partial \Theta} &  r^{(1)} \vspace{5pt}\\
\displaystyle r^{(0)} \frac{\partial \theta}{\partial R} & \displaystyle  \frac{r^{(0)}}{R}\frac{\partial \theta}{\partial \Theta} & r^{(0)} \theta^{(1)} \vspace{5pt}\\ 
\displaystyle\frac{\partial z^{(0)}}{\partial R} & \displaystyle \frac{1}{R}\frac{\partial z^{(0)}}{\partial \Theta} &  z^{(1)}
\end{bmatrix},\quad
J_G=\text{det}({\mathbf{G}})=\lambda^2. \label{rad_circum_kinematics}
\end{align}
\changesg{On substituting \eqref{rad_circum_kinematics} }in the governing equation \eqref{expand_avg_Piola}, we obtain
\begin{align}
2C_0\nabla \cdot \lambda^2 \begin{bmatrix}
\displaystyle\frac{\partial r^{(0)}}{\partial R} & \displaystyle \frac{1}{R}\frac{\partial r^{(0)}}{\partial \Theta} &  r^{(1)} \vspace{5pt}\\
\displaystyle r^{(0)} \frac{\partial \theta}{\partial R} & \displaystyle  \frac{r^{(0)}}{R}\frac{\partial \theta}{\partial \Theta} & r^{(0)} \theta^{(1)} \vspace{5pt}\\ 
\displaystyle\frac{\partial z^{(0)}}{\partial R} & \displaystyle \frac{1}{R}\frac{\partial z^{(0)}}{\partial \Theta} &  z^{(1)}
\end{bmatrix} \begin{bmatrix}
\displaystyle \frac{1}{\lambda^2} & 0 & 0\\
0 & \displaystyle \frac{1}{\lambda^2} & 0\\
0 & 0 & 1 \end{bmatrix} - \nabla \cdot \left[ p^{(0)} \lambda^2 \frac{\text{cofac}(\mathbf{F}^{(0)})}{\det \mathbf{F}^{(0)}} \right] \nonumber \\
+ ~O(h)=-\bar{\mathbf{q}},\label{rad_circum_div_eq_Piola}
\end{align}
where the unknown variables in this case are calculated as
\begin{align*}
p^{(0)}=\frac{2C_0\lambda^4}{\left|\nabla {\mathbf{x}^{(0)}}^{**}\right|^2},\quad
r^{(1)}= \displaystyle \frac{p^{(0)} \nabla x_{11}}{2 C_0 \lambda^2}, \quad
\theta^{(1)}=\frac{p^{(0)} \nabla x_{22}}{2 C_0 \lambda^2 r^{(0)}},\quad
z^{(1)}=\frac{p^{(0)} \nabla x_{33}}{2 C_0 \lambda^2}.
\end{align*}
The explicit expressions for $r^{(1)}, \theta^{(1)}, ~z^{(1)},~p^{(0)}$ are given in the \hyperref[Appendix_E] {Appendix \ref{Appendix_E}}.
Here, again we use the Winkler support on the bottom surface of the plate with traction given by \eqref{trac_Z}.
\subsubsection{Linear buckling analysis for combined growth model}
We conduct the linear bifurcation analysis by perturbing the principal solution with small parameter ($\epsilon$) to determine the onset of instability by assuming the form
\begin{align}
{r}^{(0)}({R},\Theta) &= {R} + \epsilon \Delta U(R) \cos(m \Theta),\nonumber\\
\theta &= \Theta ,\label{ansatz2}\\
z^{(0)}(R,\Theta) &= -2 h [\lambda^2-1] + \epsilon \Delta W(R) \cos(m\Theta). \nonumber
\end{align}
where $m=1,2,3...$ represents the mode number in the circumferential direction. \changes{ On substituting \eqref{ansatz2} in \eqref{rad_circum_div_eq_Piola}, making use of \eqref{mod_gov_plate_eq} along with \eqref{trac_Z} which have the updated traction component as $q_3^{+} = -K_0 \lambda^2 W_0$ and collecting only $ O(\epsilon)$ terms, we obtain the coupled differential equations for dimensionless displacement functions $U$ and $W$ as}
\changes{
\begin{align}
&2[1 + 3\lambda^6] U'' + \frac{2}{\rho} [1 + 5\lambda^6] U' - \frac{2}{\rho^2} [m^2] U \nonumber\\
&-\bar{h}\bigg[4 \lambda^2 [1 + \lambda^6] W''' + \frac{2}{\rho} \lambda^2 \big[ 1 + 5 \lambda^6 \big] W''+ \frac{2}{\rho^2} m^2 \lambda^2 [\lambda^6 - 2] W'\nonumber\\
& - \frac{2}{\rho^3} \lambda^2 m^2 [\lambda^6 - 1] W \bigg] = 0,\\
& 2 [1 - \lambda^6] W'' + \frac{2}{\rho} \left[1 - \lambda^6 \right] W' - \frac{2}{\rho^2} m^2 [1 - \lambda^6]W - \bar{h} \Bigg[ \frac{4}{\rho} \bigg[4 \lambda^8 - 4 \lambda^6 +  \lambda^2 - 1 \bigg] U'' \nonumber \\
& - \frac{2}{\rho^2} \bigg[2 m^2 \lambda^8 - 3 m^2 \lambda^6 - 2 \lambda^8 + 2 m^2 \lambda^2 - m^2 + 2 \lambda^2 \bigg] U' \nonumber\\
& - \frac{2}{\rho^3} \bigg[ \lambda^2 \big[ 2 m^2 \lambda^6 - 3 m^2 \lambda^4 + 2 \lambda^6 + m^2 -2 \big] \bigg] U \Bigg] \nonumber\\
& + \frac{2}{3} \bar{h}^2 \lambda^2 \Bigg[ -2 [1 + \lambda^6] W^{iv} - \frac{1}{\rho} [3 + 7 \lambda^6] W''' + \frac{m^2}{\rho^2} \bigg[3 + \lambda^6 \bigg]W'' \nonumber\\
& +  \frac{3 m^2}{\rho^3} \bigg[2 \lambda^6 -1 \bigg] W' + \frac{m^2}{\rho^4} \bigg[\lambda^6 m^2 - 3 \lambda^6 - m^2 + 3 \bigg]W \Bigg] \nonumber \\
&  -\frac{\beta \lambda^2}{2 \bar{h}}W + \beta \lambda^4 + \frac{\beta \lambda^4}{\rho} U - \frac{\bar{h}}{\rho} \beta [2 \lambda^6  - 1]-  \bar{h} \beta \lambda^6 W'' \nonumber\\
& -\frac{1}{3} h^2 \lambda^2 \beta \bigg[\frac{1}{\rho} [7 + 6 \lambda^6] U'' - \frac{1}{\rho^2} [9 \lambda^6 + 2 m^2 - 6] U' + \frac{1}{\rho^3} [\lambda^6 + 3 m^2 - 4] U\bigg] = 0.
\end{align}
}

\subsubsection{ \changes{Results and discussion}}
\changes{The compound matrix method is applied to solve the} governing equations to determine variation of the critical growth factor ($\lambda_{cr}$) with respect to normalized thickness ($\bar{h}$). 
\begin{figure}
\centering
\includegraphics[width=0.4\linewidth]{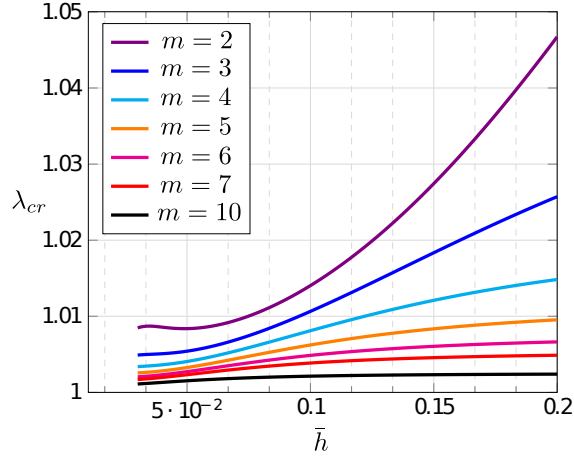}
\caption{\changes{Variation of the critical growth parameter $\lambda_{cr}$ with normalised plate thickness 
$\bar{h}$ and $\beta=0.2$ for different circumferential mode number ($m$).}}\label{h_and_L_variation_for_r_and_c_growth} 
\end{figure}
\changesg{The variation of critical buckling parameter ($\lambda_{cr}$) for different circumferential mode numbers are given in \hyperref[h_and_L_variation_for_r_and_c_growth]
{Figure \ref{h_and_L_variation_for_r_and_c_growth}}. The results shows that the higher modes arise for lower value of $\lambda_{cr}$ for all thicknesses. Also, the phenomenon of mode jump is not observed here as there is no intersection of the bifurcation curves corresponding to different mode numbers.} Furthermore, we note that the mode numbers in the case of combined radial/circumferential growth appear explicitly in the governing equations as opposed to interpretation of mode numbers in the purely radial growth case.

\begin{figure}
\centering
\includegraphics[width=0.7\linewidth]{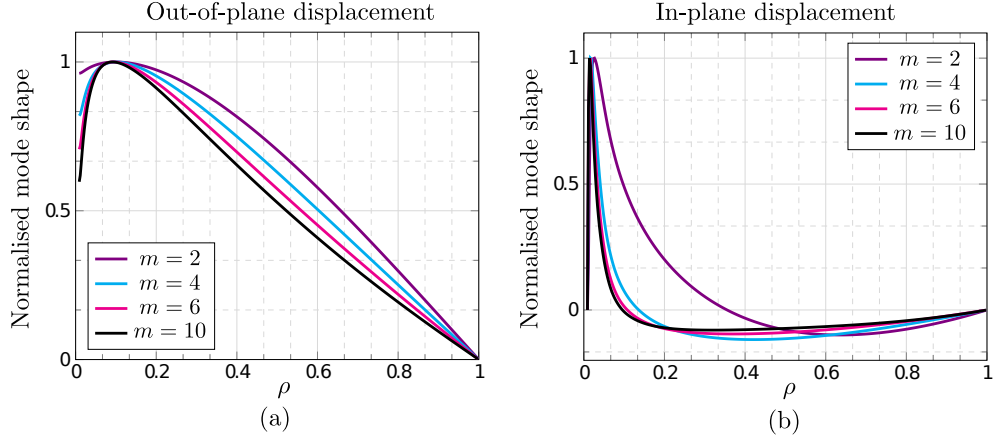}
\caption{\changes{Normalised bifurcation mode shapes for circular plate under combined radial and circumferential growth (a) {Normalised} out-of-plane displacement $W$ for mode numbers \changes{$m=2,4,6,10$ at $\bar{h} = 0.1$}. (b) {Normalised} in-plane displacement $U$ for mode number $m=2,4,6,10$ at $\bar{h} = 0.1$.}
} \label{fig_8}
\end{figure}

\noindent The normalized out of plane and in-plane displacement amplitude variation in the radial direction corresponding to each circumferential mode number is shown in \hyperref[fig_8] {Figure \ref{fig_8}} for a plate thickness of \changes{$\bar{h}=0.1$}. 
For this plate thickness, \changes{the bifurcation curve corresponding to $m=10$ mode is more stable} as we conclude by analysing \hyperref[h_and_L_variation_for_r_and_c_growth]{Figure \ref{h_and_L_variation_for_r_and_c_growth}}.
Other mode shapes may not be achievable but are plotted for completeness.

\changes{Analysing the mechanics of a circular plate under radial growth and combined growth shows that the buckling configuration of the plate changes with the increase of thickness in both the cases. The higher modes are more stable in thin regime of the plate in the former case and the bifurcation solution corresponding to higher modes are more stable for latter combined growth case regardless of the thickness. However, the critical value of bifurcation parameter ($\lambda_{cr}$) is low in combined growth case for higher modes as compared to the purely radial growth case.} 

\section{Concluding remarks} \label{section_5}
Mechanical instabilities are often observed in thin films, elastic structures and in soft biological tissues. In this paper, we have used a consistent finite-strain plate theory to investigate the buckling behaviour of incompressible circular hyperelastic plate subjected to growth. A 3-D governing system of PDEs is converted into 2-D plate system using series expansion in terms of the thickness variable. 
We consider two examples of growth-induced instability in neo-Hookean circular plate under a) radial growth and b) combined radial/circumferential growth conditions. In both the cases, the circular plate is resting on a Winkler support and subjected to simply supported boundary conditions. 
The compound matrix method is used to evaluate the critical buckling growth factor of the hyperelastic plate.
The numerical performance of compound matrix method is validated by comparing the results with the existing analytical solution for rectangular plate under uniform growth.

The variation of critical growth factor $( \lambda_{cr})$ with normalised thickness $(\bar{h})$ for both the cases is \changes{evaluated numerically and the results show that 
the plate buckles in different modes as we increase the thickness. In thin regime of the plate, higher modes are more stable and in thick regime, lower modes are more stable for purely radial growth case. However, in combined growth case the higher modes are more stable independent of the thickness of the plate.} 

The current work can be applied in modelling of thin soft biological tissues such as skin wrinkling during wound healing and ageing, modelling of wrinkling patterns in thin deployable space structures and stretchable electronics. 
Constitutive model considered in this paper is isotropic and biological materials are generally anisotropic heterogeneous materials. 
An extension of the current analysis to account for more generally applicable material models will be taken up as a future study.
We have also restricted ourselves to determine the critical buckling load, but a post-buckling analysis may provide insights on the magnitude of the out of plane deformation and the associated mode-switching response. 
These avenues are currently under investigation and we will report our results in a suitable forum at a later stage.
\section*{Acknowledgements} 
Prashant Saxena acknowledges the support of startup funds from the James Watt School of Engineering at the University of Glasgow.

\addcontentsline{toc}{section}{References}
\bibliographystyle{author-year-prashant}
\bibliography{annual_ref} 
\newpage
\begin{appendix}
\section{Appendix: Expression for Piola stress} \label{Appendix_C}
The average stress is given as
\begin{align}
\bar{\mathbf{P}}&=\frac{1}{2h}\int_{0}^{2h}\mathbf{P}dZ=\frac{1}{2h}\int_{0}^{2h}\left[\mathbf{P}^{(0)}+Z\mathbf{P}^{(1)}+\frac{Z^2}{2!}\mathbf{P}^{(2)}+\frac{Z^3}{3!}\mathbf{P}^{(3)}+O(Z^4)\right]dZ,\nonumber \\
\bar{\mathbf{P}}&=\mathbf{P}^{(0)}+h\mathbf{P}^{(1)}+\frac{2}{3}h^2\mathbf{P}^{(3)}+O(h^3).\label{C_1}
\end{align}
Subtracting the top and bottom surface traction condition using \eqref{bottom_trac} and \eqref{top_trac}
\begin{align}
\left.\mathbf{P}\mathbf{k}\right|_{Z=2h}-\left.\mathbf{P}\mathbf{k}\right|_{Z=0}=& \mathbf{P}^{(0)}\mathbf{k}+2h \mathbf{P}^{(1)}\mathbf{k}+2h^2 \mathbf{P}^{(2)} \mathbf{k}+\frac{4}{3}h^3 \mathbf{P}^{(3)}\mathbf{k}-\mathbf{P}^{(0)} \mathbf{k} + O(h^3)=\mathbf{q}^+ +\mathbf{q}^{-},\nonumber \\
=& \mathbf{P}^{(1)} \mathbf{k}+h \mathbf{P}^{(2)} \mathbf{k}+\frac{2}{3}h^2 \mathbf{P}^{(3)}\mathbf{k}=\frac{\mathbf{q}^++\mathbf{q}^{-}}{2h} + O(h^3)=\bar{\mathbf{q}}.\label{C_2}
\end{align}
Using equilibrium equation
\begin{align}
&\nabla \cdot \bar{\mathbf{P}}+\frac{\partial }{\partial Z}\left[\bar{\mathbf{P}}\mathbf{k}\right]=\mathbf{0},\label{C_3}
\end{align}
where
\begin{align*}
\frac{\partial }{\partial Z}\left[\bar{\mathbf{P}}\mathbf{k}\right]&=\frac{\partial }{\partial Z}\left[\mathbf{P}^{(0)}\mathbf{k}+h \mathbf{P}^{(1)}\mathbf{k}+\frac{2}{3}h^2 \mathbf{P}^{(3)}\mathbf{k}+O(h^3)\right]=\mathbf{P}^{(1)}\mathbf{k}+h \mathbf{P}^{(2)}\mathbf{k}+\frac{2}{3}h^2\mathbf{P}^{(3)}\mathbf{k}+O(h^3).
\end{align*}
Now, the Piola stress $(\mathbf{P})$ is described in terms of strain energy function $\left(\phi(\mathbf{F,G})=J_G \phi_0(\mathbf{A})\right)$ as
{
\begin{equation}
\begin{aligned}
\mathbf{P}&=\frac{\partial \phi(\mathbf{F,G})}{\partial \mathbf{F}}-p\frac{\partial {L}(\mathbf{F,G})}{\partial \mathbf{F}}=J_G \bigg[\frac{\partial \phi_0(\mathbf{A})}{\partial \mathbf{F}}-p\frac{\partial {L_0}(\mathbf{A})}{\partial \mathbf{F}}\bigg]\\
&=J_G \bigg[\frac{\partial \phi_{0}}{\partial \mathbf{A}}~\frac{\partial \mathbf{A}}{\partial \mathbf{F}}-p \frac{\partial L(\mathbf{A})}{\partial\mathbf{A}} \frac{\partial \mathbf{A}}{\partial \mathbf{F}} \bigg] = J_G \bigg[\frac{\partial \phi_{0}}{\partial {A}_{ij}}\frac{\partial [{FG}^{-1}]_{ij}}{\partial {F}_{lm}} -\frac{\partial \det(\mathbf{A})}{\partial {A}_{ij}} \frac{\partial [{FG}^{-1}]_{ij}}{\partial {F}_{lm}}\bigg] \\ 
&=J_G \bigg[\frac{\partial \phi_{0}}{\partial {A}_{ij}}~\frac{\partial {F}_{ik}}{\partial {F}_{lm}}{G}^{-1}_{kj}- p \det(\mathbf{A}) A_{ij}^{-T} ~ \frac{\partial {F}_{ik}}{\partial {F}_{lm}}{G}^{-1}_{kj} \bigg] ,\\
&=J_G \bigg[ \frac{\partial \phi_{0}}{\partial {A}_{ij}}~\delta_{il}\delta_{km}\mathbf{G}^{-1}_{kj} - p A_{ij}^{-T} ~ \frac{\partial {F}_{ik}}{\partial {F}_{lm}}{G}^{-1}_{kj} \bigg] = J_G \bigg[\frac{\partial \phi_{0}}{\partial {A}_{ij}}~\delta_{il} {G}^{-1}_{mj} - p A_{ij}^{-T} ~ \delta_{il} {G}^{-1}_{mj} \bigg]\\
&=\bigg[J_G\left[\frac{\partial \phi_{0}(\mathbf{A})}{\partial {A}}\right]_{lj}~{G}^{-1}_{mj} - p A_{lj}^{-T} {G}^{-1}_{mj} \bigg] = J_G \bigg[\frac{\partial \phi_{0}(\mathbf{A})}{\partial \mathbf{A}}~\mathbf{G}^{-T}-p \mathbf{A}^{-T} \mathbf{G}^{-T} \bigg], \\
&=J_G \left[\frac{\partial \phi_{0}(\mathbf {A})}{\partial \mathbf{A}}-p \mathbf{A}^{-T}\right]\mathbf{G}^{-T}. \label{C_4}
\end{aligned}
\end{equation}}

The series expansion for Piola Kirchhoff stress ($\mathbf{P}$) is given as
\begin{align}
\mathbf{P}(\mathbf{x},p)&=\mathbf{P}^{(0)}(\mathbf{x},p)+Z \mathbf{P}^{(1)}(\mathbf{x},p)+\frac{Z^2}{2} \mathbf{P}^{(2)}(\mathbf{x},p)+\frac{Z^3}{3!}\mathbf{P}^{(3)}(\mathbf{x},p)+O(Z^4).\label{C_5}
\end{align}
Using \eqref{Taylor_G} and \eqref{C_5} we have
\begin{align}
\mathbf{P}(\mathbf{x},p)=J_G\left[\frac{\partial \phi_0(\mathbf{A})}{\partial \mathbf{A}}-p\frac{\partial L_0 (\mathbf{A})}{\partial \mathbf{A}}\right]\left[\bar{\mathbf{G}}^{(0)}(\mathbf{\zeta})+Z\bar{\mathbf{G}}^{(1)}(\mathbf{\zeta})+\frac{Z^2}{2} \bar{\mathbf{G}}^{(2)}(\mathbf{\zeta})+\frac{Z^3}{3!}\bar{\mathbf{G}}^{(3)}(\mathbf{\zeta})+O(Z^4)\right].\label{C_6}
\end{align}
Substituting expansion of $\phi_0(\mathbf{A}) ~\text{and}~ L_0(\mathbf{A})$ about $\mathbf{A}^{(0)}$  we obtain
\begin{equation} 
\begin{aligned}
\mathbf{P}(\mathbf{x},p)&=J_G\bigg[\frac{\partial}{\partial \mathbf{A}}\left[\phi_0(\mathbf{A})+\frac{\partial \phi_0}{\partial \mathbf{A}}[\mathbf{A}-\mathbf{A^{(0)}}]+\frac{1}{2!}\frac{\partial^2\phi_0}{\partial \mathbf{A}\partial \mathbf{A}}[\mathbf{A}-\mathbf{A^{(0)}},~\mathbf{A}-\mathbf{A^{(0)}}]\right. \\
&\left. +\frac{1}{3!}\frac{\partial^3 \phi_0}{\partial \mathbf{A}\partial \mathbf{A}\partial \mathbf{A}}[\mathbf{A}-\mathbf{A^{(0)}},~\mathbf{A} 
-\mathbf{A^{(0)}},~\mathbf{A}-\mathbf{A^{(0)}}\right]
 \\
&-\left[p^{(0)}+Z p^{(1)}+\frac{Z^2}{2} p^{(2)}+\frac{Z^3}{3!}p^{(3)}+\frac{Z^4}{4!}p^{(4)}+O(Z^4)\right]\times \\
& \hspace{0.15in}\bigg[\frac{\partial }{\partial \mathbf{A}}\left[L_0(\mathbf{A})+\frac{\partial L_0}{\partial \mathbf{A}}[\mathbf{A}-\mathbf{A^{(0)}}]+\frac{1}{2!}\frac{\partial^2 L_0}{\partial \mathbf{A}\partial \mathbf{A}}[\mathbf{A}-\mathbf{A^{(0)}},~\mathbf{A}-\mathbf{A^{(0)}}] \right.   \\ 
 & \left.+\frac{1}{3!}\frac{\partial^3 L_0}{\partial \mathbf{A}\partial \mathbf{A}\partial \mathbf{A}}[\mathbf{A}-\mathbf{A^{(0)}},~\mathbf{A}-\mathbf{A^{(0)}},~\mathbf{A}-\mathbf{A^{(0)}}]\right]\bigg]  \\
 & \hspace{0.1in} \left[\bar{\mathbf{G}}^{(0)}(\mathbf{\zeta})+Z\bar{\mathbf{G}}^{(1)}(\mathbf{\zeta})+\frac{Z^2}{2} \bar{\mathbf{G}}^{(2)}(\mathbf{\zeta})+\frac{Z^3}{3!}\bar{\mathbf{G}}^{(3)}(\mathbf{\zeta})+O(Z^4)\right].\label{C_7}
\end{aligned}
\end{equation}
Substituting expression for $[\mathbf{A}-\mathbf{A^{(0)}}]$ using \eqref{Taylor_A} and compare the terms
\begin{equation}
\begin{aligned}
\mathbf{P}^{(0)}(\mathbf{x},p)&=J_G\left[\pmb{\mathcal{A}}^{(0)}-p^{(0)} \pmb{\mathcal{L}}^{(0)}\right]\bar{\mathbf{G}}^{(0)},\\
\mathbf{P}^{(1)}(\mathbf{x},p)&=J_G \left[ \left[ \pmb{\mathcal{A}}^{(1)}[\mathbf{A}^{(1)}]-p^{(0)} \pmb{\mathcal{L}}^{(1)}[\mathbf{A}^{(1)}]-p^{(1)} \pmb{\mathcal{L}}^{(0)}\right]\bar{\mathbf{G}}^{(0)}+\left[ \pmb{\mathcal{A}}^{(0)}-p^{(0)} \pmb{\mathcal{L}}^{(0)}\right]\bar{\mathbf{G}}^{(1)}\right],\\
\mathbf{P}^{(2)}(\mathbf{x},p)&=J_G\bigg[\bigg[\pmb{\mathcal{A}}^{(1)}[\mathbf{A}^{(2)}]+ \pmb{\mathcal{A}}^{(2)}[\mathbf{A}^{(1)},~\mathbf{A}^{(1)}]-2p^{(1)} \pmb{\mathcal{L}}^{(1)}[\mathbf{A}^{(1)}]-p^{(0)}\pmb{\mathcal{L}}^{(1)}[\mathbf{A}^{(2)}]\\
& \hspace{0.2in} -p^{(0)} \pmb{\mathcal{L}}^{(2)}[\mathbf{A}^{(1)},\mathbf{A}^{(1)}]-p^{(2)} \pmb{\mathcal{L}}^{(0)}\bigg]\bar{\mathbf{G}}^{(0)}\\
&\hspace{0.24in} +\bigg[2 \pmb{\mathcal{A}}^{(1)}[\mathbf{A}^{(1)}]-2p^{(0)}\pmb{\mathcal{L}}^{(1)}[\mathbf{A}^{(1)}] -2p^{(1)} \pmb{\mathcal{L}}^{(0)}\bigg]\bar{\mathbf{G}}^{(1)} 
 +\left[ \pmb{\mathcal{A}}^{(0)}-p^{(0)} \pmb{\mathcal{L}}^{(0)}\right]\bar{\mathbf{G}}^{(2)}\bigg].\label{C_8}
\end{aligned}
\end{equation}
where $\displaystyle \pmb{\mathcal{A}}^{{i}}(\mathbf{A}^{(0)})= \left.\frac{\partial^{i+1} \phi_0(\mathbf{A})}{\partial \mathbf{A}^{i+1}}\right|_{\mathbf{A}=\mathbf{A}^{(0)}}$ ~and~ $\displaystyle  \pmb{\mathcal{L}}^{{i}}(\mathbf{A}^{(0)})=\left.\frac{\partial R_0(\mathbf{A})}{\partial \mathbf{A}^{i+1}}\right|_{\mathbf{A}=\mathbf{A}^{(0)}}$

\section{Appendix: Expression for $\mathcal{L}^{i}$} \label{Appendix_B}
\begin{align*}
\pmb{\mathcal{L}}^{(0)}(\mathbf{A})&=\frac{\partial \det (\mathbf{A})}{\partial \mathbf{A}}= \det(\mathbf{A}) \mathbf{A}^{-T} = \det(\mathbf{A}) A_{ab}^{-T} \mathbf{e}_a \otimes \mathbf{e}_b,\\
\pmb{\mathcal{L}}^{(1)}(\mathbf{A})&=\frac{\partial^2 \det (\mathbf{A})}{\partial \mathbf{A}~\partial \mathbf{A}}=\text{det}(\mathbf{A})\bigg[\mathbf{A}^{-T} \otimes\mathbf{A}^{-T}\bigg] + \text{det}(\mathbf{A})\bigg[\mathbb{T} \changes{ [-\mathbf{A}^{-1}\boxtimes \mathbf{A}^{-T}]}\bigg],\\
&=\det(\mathbf{A}) \bigg[A_{ab}^{-T} A_{cd}^{-T} - A_{bc}^{-1} A_{ad}^{-T}\bigg] 
 \mathbf{e}_a \otimes \mathbf{e}_b \otimes \mathbf{e}_c \otimes \mathbf{e}_d,\\
\pmb{\mathcal{L}}^{(2)}(\mathbf{A})&=\frac{\partial^3 \det (\mathbf{A})}{\partial \mathbf{A}~\partial \mathbf{A}~\partial \mathbf{A}} \nonumber\\
&=\text{det}(\mathbf{A})\bigg[A_{ab}^{-T}A_{cd}^{-T}A_{ef}^{-T} - A_{cb}^{-T}A_{ad}^{-T}A_{ef}^{-T} - A^{-T}_{eb}A^{-T}_{af}A^{-T}_{cd}\\ 
& \hspace{0.2in}- A^{-T}_{ab}A^{-T}_{ed}A^{-T}_{cf} + A^{-T}_{eb}A^{-T}_{cf}A^{-T}_{ad} + A^{-T}_{cb}A^{-T}_{ed}A^{-T}_{af}\bigg]
\mathbf{e}_a\otimes \mathbf{e}_b \otimes \mathbf{e}_c \otimes \mathbf{e}_d\otimes \mathbf{e}_e \otimes \mathbf{e}_f.
\end{align*}
where,
\begin{align*}
[\mathbf{A} \otimes \mathbf{B}]_{abcd}&= \big[A_{ab} A_{cd} \big] \mathbf{e}_a \otimes \mathbf{e}_b \otimes \mathbf{e}_c \otimes \mathbf{e}_d, \\
[\mathbf{A} \boxtimes \mathbf{B}]_{abcd}&= \big[A_{ac} A_{bd} \big] \mathbf{e}_a \otimes \mathbf{e}_b \otimes \mathbf{e}_c \otimes \mathbf{e}_d,\\
[\changes{\mathbb{T} [\mathbf{A} \boxtimes \mathbf{B}]}]_{abcd}&=[A_{bc} A_{ad}] \mathbf{e}_a \otimes \mathbf{e}_b \otimes \mathbf{e}_c \otimes \mathbf{e}_d.
\end{align*}
and $\mathbb{T}_{ijkl}=\delta_{il} \delta_{jk}$.\\
Note: In this current work, we assume incompressible material thus, $\det (\mathbf{A})=1$ 

\section{Appendix: Expression for $\mathbf{x}^{(1)}$ and $p^{(0)}$} \label{Appendix_D}
For an incompressible neo-Hookean material elastic strain energy function is given as
\begin{align}
\phi_0(\mathbf{A})&=C_0[I_1-3], \nonumber\\
\phi_0(\mathbf{A})&=C_0[\text{tr}(\mathbf{A}^T\mathbf{A})-3],\label{D_1}
\end{align}
where $I_1=\text{tr}(\mathbf{C})=\text{tr}(\mathbf{A}^{T}\mathbf{A})$ as elastic strain energy depends only on elastic deformation tensor $(\mathbf{A})$ and $tr(\cdot)$ is trace of a tensor ($\cdot$). Using \eqref{C_4}, the Piola stress $(\mathbf{P})$ for incompressible neo-Hookean material is 
\begin{align}
\mathbf{P}=J_G \bigg[2C_0[\mathbf{A}]-p \mathbf{A}^{-T}\bigg]\mathbf{G}^{-T}, \label{D_2}
\end{align}
where
\begin{align*}
\frac{\partial \phi_{0}(\mathbf{A})}{\partial \mathbf{A}}&=C_0\left[\frac{\partial }{\partial \mathbf{A}}\bigg[\text{tr}(\mathbf{A}^T\mathbf{A})-3\bigg]\right]=C_0\left[\frac{\partial }{\partial {A}_{lm}}({A}^T {A})_{ii})\right]=C_0\left[\frac{\partial }{\partial {A}_{lm}}({A}_{ij} A_{ij})\right], \\
&= C_0\bigg[\delta_{il}\delta_{jm}{A}_{ij}+{A}_{ij}\delta_{il}\delta_{jm}\bigg]
=2C_0\bigg[{A}_{ij}\delta_{il}\delta_{jm}\bigg] =2C_0~{A}_{lm}=2C_0~\mathbf{A}. 
\end{align*}
Now, the first term $\mathbf{P}^{(0)}$ in \eqref{C_5}  for neo-Hookean material is rewritten as
\begin{align}
\mathbf{P}^{(0)}&=J_G\left[2C_0\mathbf{A}^{(0)}-p{\mathbf{A}^{(0)}}^{-T}\right]\bar{\mathbf{G}}^{(0)}. \label{D_3}
\end{align}
By using bottom traction condition, $\mathbf{P}^{(0)} \mathbf{k}=-\mathbf{q}^-$ and substituting expression for $\mathbf{A}^{(0)}$ we have
\begin{align}
\mathbf{P}^{(0)}\mathbf{k}&=2C_0J_G\mathbf{F}^{(0)}\bar{\mathbf{G}}^{{(0)}^{T}}  \bar{\mathbf{G}}^{(0)}\mathbf{k}-J_G ~ p^{(0)}\mathbf{F}^{{(0)}^{-T}}\bar{\mathbf{G}}^{{(0)}^{-1}}\bar{\mathbf{G}}^{(0)}\mathbf{k},\nonumber\\
-\mathbf{q}^{-}&=2C_0 \bigg[\nabla \mathbf{x}^{(0)}\bar{\mathbf{G}}^{{(0)}^{T}}+\mathbf{x}^{(1)}\otimes \bar{\mathbf{G}}^{(0)}\mathbf{k}\widehat{\mathbf{G}}^{(0)}\mathbf{k}\bigg]-J^{(0)}p^{(0)} \mathbf{F}^{{(0)}^{-T}}\mathbf{k},\nonumber\\
&=2C_0\bigg[\nabla \mathbf{x}^{(0)}\bar{\mathbf{G}}^{{(0)}^{T}}(\widehat{\mathbf{G}}^{(0)})\mathbf{k}+\left[\bar{\mathbf{G}}^{(0)}\mathbf{k}\cdot \widehat{\mathbf{G}}^{(0)} \mathbf{k}\right]\mathbf{x}^{(1)}\bigg]-\left[J^{(0)}p^{(0)} \frac{\text{Cofac}(\mathbf{F}^{(0)})}{\text{det}(\mathbf{A}^{(0)}){\text{det}(\mathbf{G}^{(0)})}}\right]\mathbf{k},\nonumber\\
&=2C_0\nabla \mathbf{x}^{(0)}\bar{\mathbf{G}}^{{(0)}^{T}} \widehat{\mathbf{G}}^{(0)} \mathbf{k}+2C_0J^{(0)}\left|\bar{\mathbf{G}}^{(0)}\mathbf{k}\right|^2\mathbf{x}^{(1)}-\left[J^{(0)}p^{(0)} \frac{\text{Cofac}(\mathbf{F}^{(0)})}{(1)J^{(0)}}\right]\mathbf{k}.\label{D_4}
\end{align}
where $J^{(0)}=\left. J_G \right|_{Z=0}$. Traction at bottom surface is given as
\begin{align}
-\mathbf{q}^{-}&=2C_0\nabla \mathbf{x}^{(0)}\bar{\mathbf{G}}^{{(0)}^{T}} \widehat{\mathbf{G}}^{(0)}\mathbf{k}+2C_0 J_G \left|\bar{\mathbf{G}}^{(0)}\mathbf{k}\right|^2\mathbf{x}^{(1)}-p^{(0)} \text{Cofac}(\mathbf{F}^{(0)})\mathbf{k}, \label{D_5}
\end{align}
where $\text{Cofac}(\mathbf{F}^{(0)})\mathbf{k}=\text{Cofac}(F^{(0)})_{ij}k_j\mathbf{e}_i$. In this work, we use  $\nabla \mathbf{x}^{{(0)}^*}$ in place of $\text{Cofac}(\mathbf{F}^{(0)})\mathbf{k}$ and for deformation gradient in polar coordinate system it is given by
\begin{align}
\nabla \mathbf{x}^{{(0)}^*}=\displaystyle  \frac{r^{(0)}}{R}\left[\frac{\partial \theta}{\partial R}\displaystyle\frac{\partial z^{(0)}}{\partial \Theta}- \displaystyle\frac{\partial \theta}{\partial \Theta}\displaystyle\frac{\partial z^{(0)}}{\partial R}\right] \mathbf{e}_1+\displaystyle \frac{1}{R}\left[\displaystyle\frac{\partial r^{(0)}}{\partial \Theta}\displaystyle\frac{\partial z^{(0)}}{\partial R}-\displaystyle\frac{\partial r^{(0)}}{\partial R}\displaystyle\frac{\partial z^{(0)}}{\partial \Theta}\right]\mathbf{e}_2 \nonumber \\
+ \displaystyle \frac{r^{(0)}}{R} \left[\displaystyle\frac{\partial r^{(0)}}{\partial R}\displaystyle\frac{\partial \theta}{\partial \Theta}-\displaystyle\frac{\partial \theta}{\partial R}\displaystyle\frac{\partial r^{(0)}}{\partial \Theta}\right]\mathbf{k}.\label{D_6}
\end{align}
We know from \eqref{incompressibility_cond}
\begin{align}
&\text{det}(\mathbf{A}^{(0)})=1 =
 \text{det}(\mathbf{F}^{(0)})\text{det}(\bar{\mathbf{G}}^{{(0)}^{T}}),\nonumber\\
&\text{det}(\mathbf{F}^{(0)})=\text{det}(\bar{\mathbf{G}}^{{(0)}^{-T}}).  \label{D_7}
\end{align}
As the definition of determinant
\begin{align}
\text{det}({\mathbf{F}}^{(0)})&=\bigg[r^{(1)}\mathbf{e}_1 + r^{(0)} \theta^{(1)} \mathbf{e}_2 + z^{(1)} \mathbf{e}_3 \bigg] \cdot \text{Cofac}(\mathbf{F}^{(0)}) \mathbf{k}= \mathbf{x}^{(1)} \cdot \nabla \mathbf{x}^{{(0)}^*}.\label{D_8}
\end{align}
By combining \eqref{D_7} and \eqref{D_8} we have
\begin{align}
\mathbf{x}^{(1)} \cdot \nabla \mathbf{x}^{{(0)}^*}=\det\left(\bar{\mathbf{G}}^{{(0)}^{-T}}\right). \label{D_9}
\end{align}
Using \eqref{D_5} we obtain the explicit expression for $\mathbf{x}^{(1)}$ 
\begin{align}
\mathbf{x}^{(1)}&=\frac{-\mathbf{q}^{-}-2C_0\nabla \mathbf{x}^{(0)}\bar{\mathbf{G}}^{{(0)}^{T}} \widehat{\mathbf{G}}^{(0)}\mathbf{k}+p^{(0)}\nabla \mathbf{x}^{{(0)}^*}}{2C_0 J_G \left|\bar{\mathbf{G}}^{(0)}\mathbf{k}\right|^2}.\label{D_10}
\end{align}
To obtain the explicit expression for $p^{(0)}$ we substitute \eqref{D_10} into \eqref{D_9} which yields 
\begin{align}
p^{(0)}=\frac{2C_0 J_G \left|\bar{\mathbf{G}}^{(0)}\mathbf{k}\right|^2}{\text{det}\bar{\mathbf{G}}^{{(0)}^{T}}\left|\nabla \mathbf{x}^{{(0)}^*}\right|^{2}}+\left[\mathbf{q}^{-} + 2C_0\nabla \mathbf{x}^{(0)}\bar{\mathbf{G}}^{{(0)}^{T}} \widehat{\mathbf{G}}^{(0)}\mathbf{k}\right] \cdot \frac{\nabla \mathbf{x}^{{(0)}^*}}{{\left|\nabla \mathbf{x}^{{(0)}^*}\right|^{2}}}.\label{D_11}
\end{align}

\section{Appendix: General Description on Compound Matrix Method } \label{Appendix_A} 
Consider a two-point boundary value problem expressed in first order ordinary differential equations
\begin{align}
\frac{d \mathbf{Y}}{dX}= \pmb{\mathscr{A}}(\lambda,x) \mathbf{Y}, \hspace{0.8in} x \in (a,b) \label{A_1}
\end{align} 
subjected to boundary condition
\begin{align}
\mathbf{BY}&=\mathbf{0}, \hspace{1in} x=a, \nonumber\\
\mathbf{CY}&=\mathbf{0}, \hspace{1in} x=b, \label{A_2}
\end{align}
where $\lambda$ is the eigenvalue or critical buckling parameter, $\mathbf{Y}$ is $1 \times 2n$ vector, $\pmb{\mathscr{A}}$ is $2n \times 2n$ matrix and $\mathbf{B}$ and $\mathbf{C}$ both are $n \times 2n $ full rank matrices i.e., $n$ boundary conditions are given at  
$x=a,b$. Assume
\begin{align}
\bigg\{\mathbf{y}^{(1)} (\lambda,x), \mathbf{y}^{(2)} (\lambda, x), ..., \mathbf{y}^{(n)}(\lambda, x)\bigg\},
\end{align}
is set of $n$ linearly independent solution \eqref{A_1} which satisfy the boundary condition at $x=0$ and the general solution of \eqref{A_1} can be written as the linear combination of its independent solution
\begin{align}
\mathbf{y}(\lambda,x)= \sum_{j=1}^{n} k_j \mathbf{y}^{j}, \label{A_4}
\end{align}
where $k_1, k_2, ...k_n$ are the constants. Solution matrix $\mathbf{M}$ to be $2n \times n$ is define whose $j$th column is $\mathbf{y}^{(j)}$ as $\mathbf{M}=[\mathbf{y}^{(1)}, \mathbf{y}^{(2)},  ... \mathbf{y}^{(n)} ]$, then \eqref{A_1} in terms of $\mathbf{M}$ is given
\begin{align}
\frac{d \mathbf M}{dx}=[\pmb{\mathscr{A}} \mathbf{y}^{(1)}, \pmb{\mathscr{A}} \mathbf{y}^{(2)}, . . .,\pmb{\mathscr{A}} \mathbf{y}^{(n)} ]= \pmb{\mathscr{A}} \mathbf{M}. \label{A_5}
\end{align}
The compound variables are defined as minors of $\mathbf{M}$ and denoted as $\Phi_1, \Phi_2, ...$ and those are $(^{2n}C_n)$ in numbers. For an instance  consider fourth order ODE ($n=2$), then the solution matrix is
\begin{align}
\mathbf{M}=\begin{bmatrix}
y_1^{(1)} & y_1^{(2)}  \\ 
y_2^{(1)} & y_2^{(2)} \\
y_3^{(1)} & y_3^{(2)} \\
y_4^{(1)} & y_4^{(2)} 
\end{bmatrix}, \label{A_6}
\end{align}
and 6 minors of $\mathbf{M}$
\begin{align}
\Phi_1&=(1,2)=  \begin{vmatrix}
y_1^{(1)} & y_1^{(2)} \\
y_2^{(1)} & y_2^{(2)}
\end{vmatrix}, \hspace{0.5 in} \Phi_2=(1,3)=  \begin{vmatrix}
y_1^{(1)} & y_1^{(2)} \\
y_3^{(1)} & y_3^{(2)}
\end{vmatrix}, \nonumber \\[3pt]
\Phi_{3}&=(1,4), ~~ \Phi_{4}=(2,3),  ~~ \Phi_{5}=(2,4), ~~ \Phi_{6}=(3,4). \label{A_7}
\end{align}
Using \eqref{A_1}, the system of first order differential equation in terms of compound variable is
\changes{ 
\begin{align}
\Phi_1'=\begin{vmatrix}
y_1^{(1)} & y_1^{(2)}  \\ 
y_2^{(1)} & y_2^{(2)}
\end{vmatrix}'&=\begin{vmatrix}
{y_1^{(1)}}' & {y_1^{(2)}}'  \\ 
{y_2^{(1)}} & {y_2^{(2)}} 
\end{vmatrix} +
\begin{vmatrix}
{y_1^{(1)}} & {y_1^{(2)}}  \\ 
{y_2^{(1)}}' & {y_2^{(2)}}' 
\end{vmatrix}, \nonumber \\
&=\begin{vmatrix}
\sum_{j=1}^{4} \mathscr{A}_{1j} y_j^{(1)} & \sum_{j=1}^{4} \mathscr{A}_{1j} y_j^{(2)} \\
{y_2^{(1)}} & {y_2^{(2)}}
\end{vmatrix} + \begin{vmatrix}
y_1^{(1)} & y_1^{(2)}  \nonumber\\
\sum_{j=1}^{4} \mathscr{A}_{2j} y_j^{(1)} & \sum_{j=1}^{4} \mathscr{A}_{2j} y_j^{(2)}
\end{vmatrix}, \nonumber\\
&= \mathscr{A}_{11} \Phi_1 - \mathscr{A}_{13} \Phi_4 - \mathscr{A}_{14} \Phi_{5} + \mathscr{A}_{22} \Phi_1 + \mathscr{A}_{23} \Phi_{2} +\mathscr{A}_{24} \Phi_3. \label{A_8}
\end{align} }
The system is now converted into $(^{2n}C_n)$ ordinary differential equations which is in the form of
\changes{
\begin{align}
\mathbf{\Phi'}= \pmb{\mathscr{A}}^{*}(\lambda,x) \mathbf{\Phi}, \hspace{1in} x \in (a,b). \label{A_9}
\end{align} } 
Subjected to initial condition at $x=0$,
\begin{align} 
\mathbf{\Phi}(a)=[\Phi_1,~ \Phi_2,~ \Phi_3,~ \Phi_4,~ \Phi_5,~ \Phi_6] \label{A_10}
\end{align} 
Now, if we consider sixth order ODE system ($n=3$) then the solution matrix is defined as
\begin{align}
\mathbf{M}=\begin{bmatrix}
y_1^{(1)} & y_1^{(2)} & y_1^{(3)} \\ 
y_2^{(1)} & y_2^{(2)} & y_2^{(3)} \\
y_3^{(1)} & y_3^{(2)} & y_3^{(3)} \\
y_4^{(1)} & y_4^{(2)} & y_4^{(3)} \\
y_5^{(1)} & y_5^{(2)} & y_5^{(3)} \\
y_6^{(1)} & y_6^{(2)} & y_6^{(3)}
\end{bmatrix},\label{A_11}
\end{align}
where 20 minors of $\mathbf{M}$ are
\begin{align*}
&\Phi_1=(1,2,3),~~~ \Phi_2=(1,2,4), ~~~\Phi_3=(1,2,5),~~~ \Phi_4=(1,2,6),~~~\Phi_5=(1,3,4),\\
& \Phi_6=(1,3,5),~~~\Phi_7=(1,3,6),~~~ \Phi_8=(1,4,5),~~~ \Phi_9=(1,4,6),~~~\Phi_{10}=(1,5,6),\\
& \Phi_{11}=(2,3,4),~~ \Phi_{12}=(2,3,5),~~\Phi_{13}=(2,3,6),~~ \Phi_{14}=(2,4,5),~~\Phi_{15}=(2,4,6),\\
 &\Phi_{16}=(2,5,6),~~ \Phi_{17}=(3,4,5),~~\Phi_{18}=(3,4,6),~~\Phi_{19}=(3,5,6),~~ \Phi_{20}=(4,5,6).
\end{align*}
The system of equation \eqref{A_9} is now numerically integrated using initial condition \eqref{A_10} which produces the solution $y^{(j)}$ at $x=b$
\begin{align}
\mathbf{Cy}=\mathbf{C} \sum_{j=1}^{n} k_j \mathbf{y}^{(j)}(b)=\mathbf{CMk}=\mathbf{0}.\label{A_12}
\end{align}
For existence of non-trivial solution of differential equation \eqref{A_9} 
\begin{align}
\det(\mathbf{CM})=0.\label{A_13}
\end{align} 

\section{Appendix: Expression for unknown variable for radial and circumferential growth} \label{Appendix_E}
{Using Eq. \eqref{D_11} we obtain the expression for $p^{(0)}$ as
\begin{align}
p^{(0)}=\frac{2C_0 \lambda^4}{\left|\nabla {\mathbf{x}^{(0)}}^{**}\right|^2},
\end{align}
where $\nabla {\mathbf{x}^{(0)}}^{**}=\nabla {x}_{11} \mathbf{e}_1 + \nabla x_{22} \mathbf{e}_2 + \nabla x_{33} \mathbf{e}_3$ and
\begin{align*}
\nabla {x}_{11} =\frac{r^{(0)}}{R}\left[\frac{\partial \theta}{\partial R}\frac{\partial z^{(0)}}{\partial \Theta}- \frac{\partial \theta}{\partial \Theta}\frac{\partial z^{(0)}}{\partial R}\right],  \nabla {x}_{22}=\frac{1}{R}\left[\frac{\partial r^{(0)}}{\partial \Theta}\frac{\partial z^{(0)}}{\partial R}-\frac{\partial r^{(0)}}{\partial R}\frac{\partial z^{(0)}}{\partial \Theta}\right],\\
 \nabla {x}_{33}=\frac{r^{(0)}}{R} \left[\frac{\partial r^{(0)}}{\partial R}\frac{\partial \theta}{\partial \Theta}-\frac{\partial \theta}{\partial R}\frac{\partial r^{(0)}}{\partial \Theta}\right].
\end{align*}
On substituting the $p^{(0)}$ in \eqref{D_10} we obtain
\begin{align}
r^{(1)}= \displaystyle \frac{p^{(0)} \nabla x_{11}}{2 C_0 \lambda^2}, \quad
\theta^{(1)}=\frac{p^{(0)} \nabla x_{22}}{2 C_0 \lambda^2 r^{(0)}}, \quad
z^{(1)}=\frac{p^{(0)} \nabla x_{33}}{2 C_0 \lambda^2}.
\end{align}} 
\end{appendix}
\end{document}